\documentclass{jaa}
\usepackage{natbib}
\usepackage{graphicx}

\begin{document}\sloppy

\title{A tale of two nearby dwarf irregular galaxies WLM and IC~2574 - as revealed by UVIT}


\author{Chayan Mondal\textsuperscript{1,2,*}, Annapurni Subramaniam\textsuperscript{1}, Koshy George\textsuperscript{3}}
\affilOne{\textsuperscript{1}Indian Institute of Astrophysics, 2nd Block, Koramangala, Bangalore 560034\\}
\affilTwo{\textsuperscript{2}Pondicherry University, R.V. Nagar, Kalapet, 605014, Puducherry, India\\}
\affilThree{\textsuperscript{3}Faculty of Physics, Ludwig-Maximilians-Universit{\"a}t, Scheinerstr. 1, Munich, 81679, Germany}


\twocolumn[{
\maketitle

\corres{chayan@iiap.res.in}

\msinfo{}{}{}

\begin{abstract}

We present an ultra-violet study of two nearby dwarf irregular galaxies WLM and IC~2574, using the Far-UV and Near-UV data from the Ultra-Violet Imaging Telescope (UVIT). We used the F148W band Far-UV images and identified 180 and 782 young star-forming clumps in WLM and IC~2574, respectively. The identified clumps have sizes between 7 - 30 pc in WLM and 26 - 150 pc in IC~2574. We noticed more prominent hierarchical splitting in the structure of star-forming regions at different flux levels in IC~2574 than WLM. We found that the majority of the clumps have elongated shapes in the sky plane with ellipticity ($\epsilon$) greater than 0.6 in both the galaxies. The major axis of the identified clumps is found to show no specific trend of orientation in IC~2574, whereas in WLM the majority are aligned along south-west to north-east direction. We estimated (F148W$-$N242W) colour for the clumps identified in WLM and noticed that the younger ones (with (F148W$-$N242W) $<-0.5$) are smaller in size ($<10$ pc) and are located mostly in the southern half of the galaxy between galactocentric radii 0.4 - 0.8 kpc.

\end{abstract}
\keywords{galaxies: dwarf irregular - galaxies: galaxies: individual - galaxies: star formation - ultra-violet}
}]
\doinum{}
\artcitid{}
\volnum{}
\year{}
\pgrange{1--}
\setcounter{page}{1}
\lp{12}

\section{Introduction}

Dwarf galaxies show wide variation in their star formation history \citep{tolstoy2009, mcquinn2015,cignoni2018}. They are typically gas-rich systems with a range of gas-to-mass ratios observed among different sub-classes. Interaction with a massive companion or even a nearby dwarf can produce episodes of starburst in dwarf galaxies and thus can sustain star formation for a longer time \citep{patton2013,stierwalt2015}. The duration of such bursts also varies among different dwarf systems. Dwarf galaxies are also unique in terms of their physical characteristics. These systems have a low gravitational potential well, metal-poor environment, and a rigid-body like rotation with no density waves. Such unique nature of dwarf galaxies provides a scope to study star formation in extreme conditions. As dwarf galaxies mostly have low metallicity, studying these systems will be important to know how star formation proceeded in the earlier epochs \citep{weisz2011}. The absence of spiral density waves and shear force in dwarfs further offers an opportunity to study the nature of star-forming clumps in a completely different environment than the massive spiral galaxies.  

With growing evidence from both observation and simulation, we understand that star formation is a hierarchical process from smaller cores to larger complexes \citep{elmegreen2000,grasha2017}. Especially, the young massive stars are found in loosely bound groups with no specific length scale \citep{garcia2010}. Studying young star-forming regions in dwarf galaxies hence is crucial to know the properties of young stellar groups and how they continue to form in such environments. Apart from the external perturbation, star formation in dwarfs is also controlled by the stochastic self-propagating mode and the internal stellar feedback \citep{gerola1980,hunter1997}. The feedback from massive stars that formed earlier can inhibit or induce secondary star formation \citep{cignoni2019}. The distribution of young star-forming regions in isolated dwarfs hence partly depends on the impact of stellar feedback across the galaxy.

As young and massive OB type stars emit mostly in the far-UV wavelength, observations in Far-UV (FUV) band can directly trace young star-forming regions in a galaxy \citep{kennicutt2012}. The (FUV$-$NUV) colour also serves as an important quantity to constrain the age of stellar clumps up to a few hundred Myr \citep{goddard2010,mondal2019jaa}. Several studies have used UV data from the Hubble Space Telescope (HST) to understand the properties of star-forming regions from the analysis of resolved stellar population \citep{bianchi2012a,bianchi2014c,calzetti2015}.
\citet{bianchi2012} used HST observations from FUV to {\it I} band to study active star-forming regions in six nearby dwarfs and highlighted the importance of UV observations. The UV space telescope Galaxy Evolution Explorer (GALEX) has made a phenomenal contribution to the study of star-forming regions in nearby galaxies with FUV and Near-UV (NUV) broad band imaging \citep{gildepaz2007,thilker2007,kang2009,melena2009,goddard2010}. \citet{melena2009} used the GALEX FUV and NUV data combined with multi-band optical {\it UBV} and infrared {\it JHK} observations of young star-forming knots in 11 dwarf galaxies to study their properties.

In this paper, we aim to understand the properties of young star-forming clumps in two nearby dwarf irregular galaxy WLM and IC~2574. The galaxy WLM, a member of the Local Group, is located at a distance of 995 kpc \citep{urbaneja2008}. It is a relatively smaller, less massive, and metal-poor gas-rich system (parameters listed in Table \ref{wlm_table}). The galaxy has an isolated location in the sky and does not show any signature of interaction \citep{leaman2012}. The other galaxy IC~2574, located at a distance of 3.79 Mpc, is relatively large, massive, and metal-rich than WLM (parameters listed in Table \ref{ic2574_table}). IC~2574 is a member of the M81 group, and it also does not have signature of interaction \citep{yun1999}. 

Both the galaxies have been studied in UV with observations from different telescopes. \citet{bianchi2012} studied WLM with HST multi-band observations including FUV and NUV and noticed active star formation during the last 10 Myr. \citet{melena2009} have performed a photometric study of star-forming knots in WLM using GALEX data. \citet{mondal2018} studied the (FUV$-$NUV) colour demographics of young star-forming regions with multi-band imaging observations from the Ultra-Violet Imaging Telescope (UVIT). The galaxy IC~2574 has been studied by \citet{mondal2019} to understand the connection between expanding H~I holes and the triggered star formation using UVIT FUV observations.

In this work, we studied the physical properties of the FUV-bright young star-forming clumps in the galaxies WLM and IC~2574 using UVIT broadband imaging observations. The UVIT PSF ($\sim$1.4$^{\prime\prime}$) could resolve star-forming clumps up to $\sim$ 7 pc in WLM and $\sim$ 26 pc in IC~2574. The earlier studies on these galaxies with UVIT have not performed quantitative measurement of the individual star-forming clump up to the scales resolvable by the telescope. Here, we used the UVIT FUV intensity map to identify star-forming clumps and compared their size, shape, orientation, FUV magnitude, and UV colour. As the two galaxies have significantly different size and mass, we further compared the properties of star-forming clumps identified in both the system. In the next section, we present the data and observations used in this study. The analysis of the work is discussed in Section \ref{s_analysis}, results and discussions in Section \ref{s_results}, followed by summary in section \ref{s_summary}

 \begin{figure*}
\begin{center}
\includegraphics[width=6in]{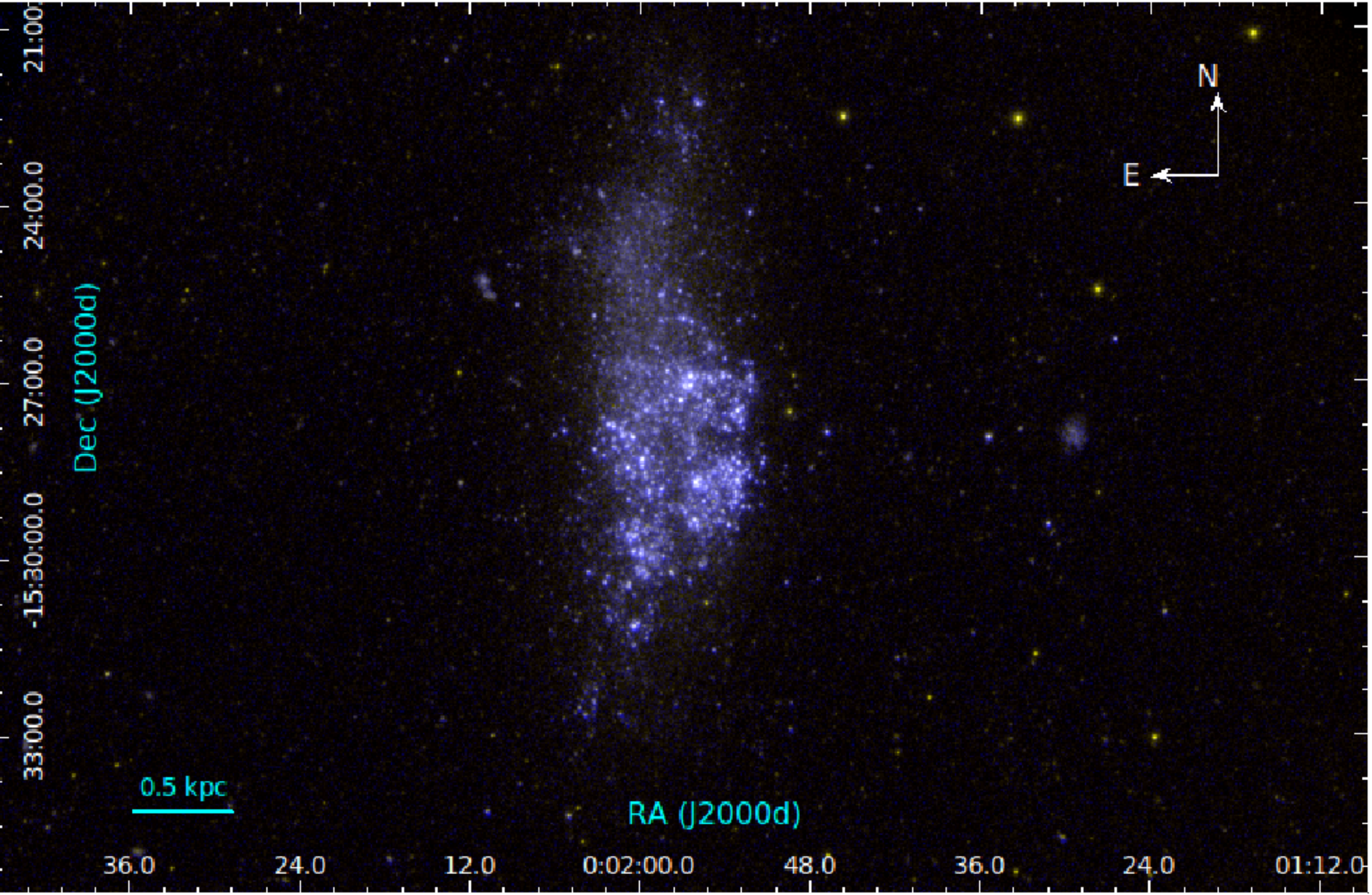} 
\caption{UVIT colour composite image of the galaxy WLM. The FUV F148W and the NUV N242W bands are shown in blue and yellow colour respectively.}
 \label{wlm_color}
 \end{center}
 \end{figure*}

 \begin{figure*}
\begin{center}
\includegraphics[width=6in]{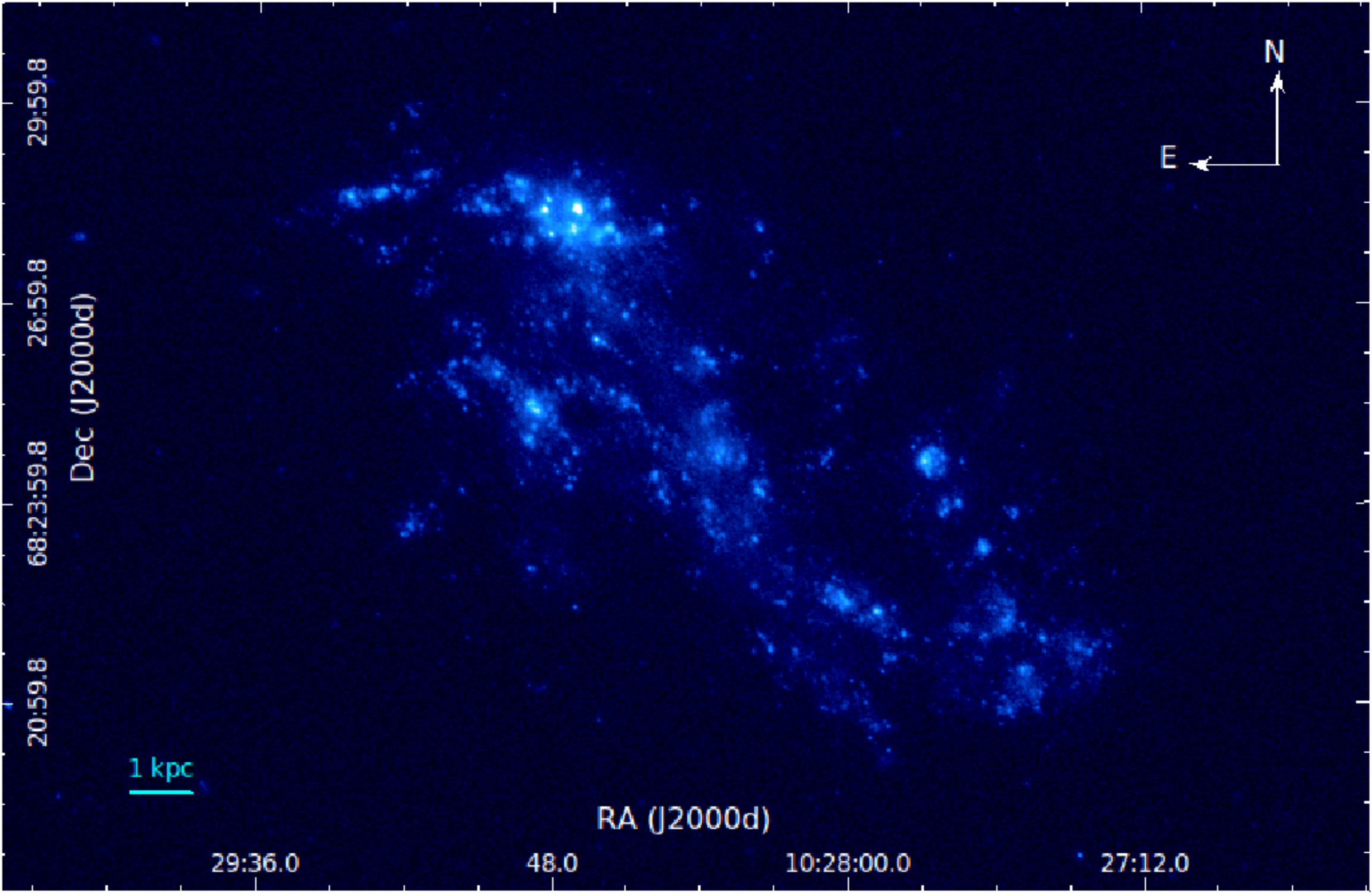} 
\caption{UVIT F148W band image of the galaxy IC~2574.}
 \label{ic2574_color}
 \end{center}
 \end{figure*}
 
\begin{table}
\centering
\caption{Properties of WLM}
 \label{wlm_table}
\resizebox{85mm}{!}{

\begin{tabular}{ccc}
\hline
 Property & Value & Reference\\\hline
 RA & 00 01 57.8 & \citet{gallouet1975}\\
 DEC & $-$15 27 51.0 & \citet{gallouet1975}\\
 Distance & 0.995 Mpc & \citet{urbaneja2008}\\
 Metallicity & 0.003 & \citet{urbaneja2008}\\
 log M$_{*} (M_{\odot})$ & 6.88 & \citet{lee2006} \\
 Major axis & 5.7$^{\prime}$ & \citet{devaucouleurs1991b}\\
 Inclination & $69^\circ$ & \citet{devaucouleurs1991b}\\
 PA of major axis & $181^\circ$ & \citet{devaucouleurs1991b}\\\hline

\end{tabular}
}
\end{table}

\begin{table}
\centering
\caption{Properties of IC 2574}
 \label{ic2574_table}
\resizebox{85mm}{!}{
\begin{tabular}{ccc}
\hline
 Property & Value & Reference\\\hline
 RA & 10 28 23.5 & \citet{skrutskie2006}\\
 DEC & +68 24 43.7 & \citet{skrutskie2006}\\
 Distance & 3.79 Mpc & \citet{dalcanton2009}\\
 Metallicity (Z) & $0.006$ & \citet{cannon2005}\\
 log M$_{*} (M_{\odot})$ & 8.39 & \citet{lee2006}\\
 Major axis & 6.7$^{\prime}$ & \citet{devaucouleurs1991b}\\
 Inclination & $63^\circ$ & \cite{pasquali2008}\\
 PA of major axis & $55^\circ$ & \citet{pasquali2008}\\\hline
\end{tabular}
}
\end{table}

\section{Data and observation}
\label{s_data}
We have used FUV and NUV imaging observations from the UVIT in this work. UVIT on AstroSat is composed of two telescopes \citep{kumar2012}. One telescope observes in the FUV waveband (1300 - 1800 $\AA{}$), whereas the other one operates in both NUV (2000 - 3000 $\AA{}$) and Visible. Both the FUV and NUV channel consists of multiple filters of different bandwidth. The observation in the Visible channel is used to correct the drift caused in the image during the course of observation. Apart from having multiple filters and a wide circular field of view of diameter 28$^{\prime}$, the UVIT instrument also has around three times better spatial resolution (FWHM of PSF $\sim$ 1.4$^{\prime\prime}$) than the Galaxy Evolution Explorer (GALEX \citet{martin2005}). Such a unique combination has provided a great advantage to study external galaxies using UVIT.

In this study, we have used F148W band FUV and N242W band NUV imaging data for the galaxy WLM (Figure \ref{wlm_color}), whereas for IC~2574, we have used the data in F148W band (Figure \ref{ic2574_color}). Each observation was performed with multiple orbits of the AstroSat satellite. The raw data acquired from the UVIT observation were processed with the help of a customised software CCDLAB to produce science ready images \citep{postma2017}. During this conversion, the images are drift-corrected, flat-fielded, aligned, and combined to produce the final deep images. We have also corrected the data for intrinsic distortion and fixed pattern noise of the detector using the calibration files \citep{girish2017,postma2011}. The final images have 4096$\times$4096 pixel dimension with 1 pixel corresponding to 0.4$^{\prime\prime}$. At the distance of the galaxy WLM and IC~2574, this single pixel corresponds to $\sim$ 2 pc and $\sim$ 7.6 pc, respectively. We have listed the exposure time of the UVIT observations and the filter combination in Table \ref{uvit_obs}. The calibration measurements are adopted from \citet{tandon2017}.

\begin{table*}
\centering
\caption{Details of UVIT bands and observations}
\label{uvit_obs}
\begin{tabular}{p{2cm}p{2cm}p{2cm}p{3.0cm}p{1.5cm}p{2cm}p{2cm}}
\hline
Filter & Bandpass ($\AA$) & ZP magnitude (AB) & Unit conversion (erg/sec/cm$^2$/$\AA$) & $\triangle \lambda (\AA)$ & Galaxy & Exposure time (sec)\\\hline
F148W & 1250-1750 & 18.016 & 3.09$\times10^{-15}$ & 500 & WLM & 2634\\
 &  &  &  &  & IC~2574 & 10375\\\hline
N242W & 2000-3000 & 19.81 & 2.22$\times10^{-16}$ & 785 & WLM & 2824\\\hline

\end{tabular}

\end{table*}

\section{Analysis}
\label{s_analysis}

\subsection{Identification of star-forming clumps}

\begin{figure}
\begin{center}
\includegraphics[width=3.2in]{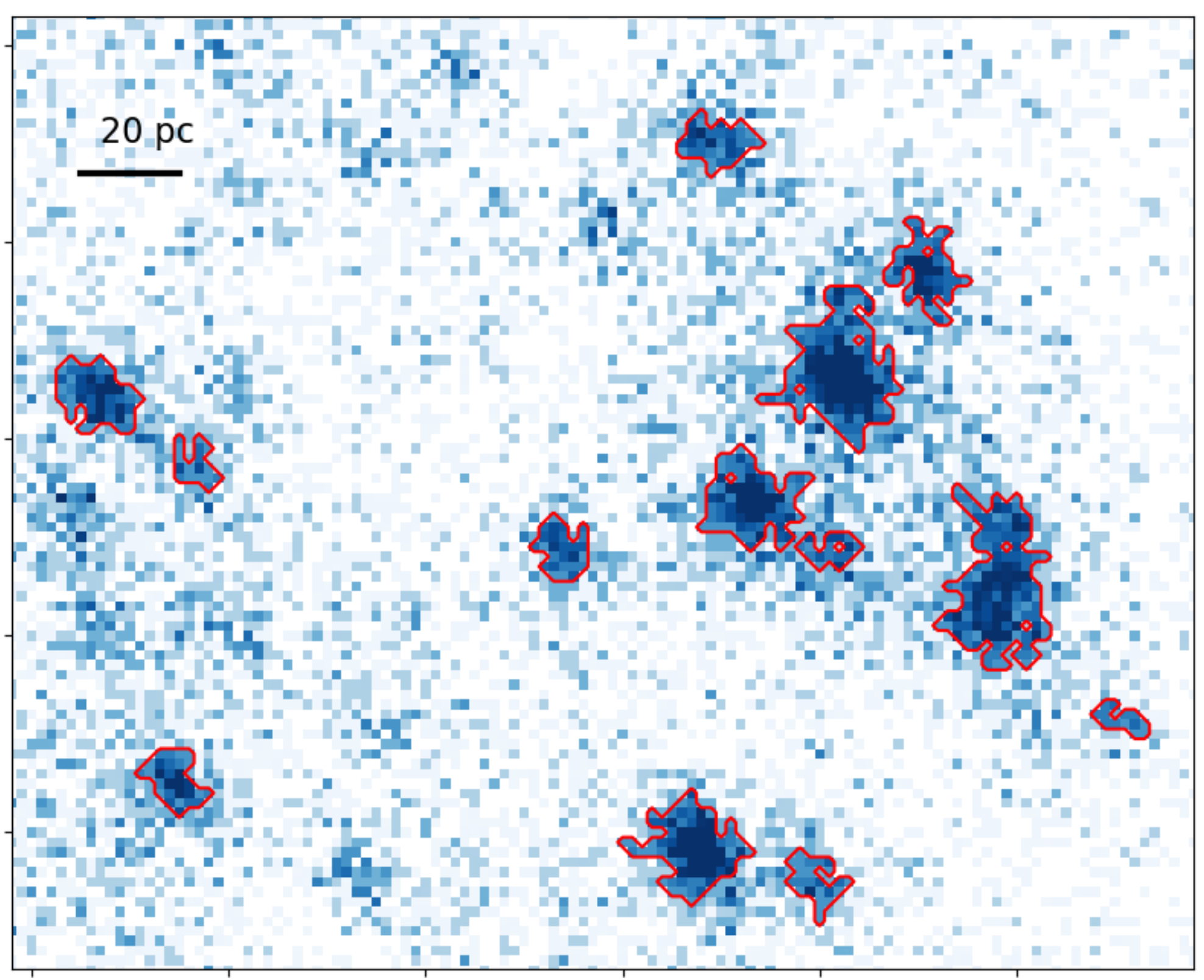} 
 \caption{A specific region of the galaxy WLM is shown to display the detection of star-forming clumps (child structure - brown contoured region) by {\it astrodendro} package.}
 \label{wlm_clump_dendro}
 \end{center}
 \end{figure}

As the young massive stars emit a copious amount of radiation in the FUV wavelength, we used UVIT F148W band images to locate the young-star forming regions in both the galaxies. The size of UVIT PSF is not sufficient to resolve individual stars in these galaxies. The point-like objects or the extended clumps detected with UVIT are actually a group of clusters, associations, or a combination of several such groups. We used F148W band FUV images and employed {\it astrodendro}\footnote{http://www.dendrograms.org/} package to identify the FUV bright star-forming clumps in both WLM and IC 2574. The {\it astrodendro} package helps to identify structures (i.e., intensity peaks) in the intensity map for a given threshold flux and a minimum clump size in pixel unit (Figure \ref{wlm_clump_dendro}). The algorithm identifies individual intensity peaks above the adopted threshold in the image and builds a structure tree (i.e., dendrogram) from the higher to lower flux levels. An intensity peak identified above the threshold is considered as a structure only if the peak height from the local minima is more than the value of minimum delta (another input parameter which we considered as 3 times the average background) and the size is greater than the defined minimum clump size. For a given set of input parameters, the code identifies both child and parent structure (in the form of a structure tree) and provides their position, area, flux, etc. The child structures are individual intensity peaks that could not be resolved further with UVIT, whereas the parent structures are defined as those that contain multiple child structures inside them. Based on the shape and intensity profile, {\it astrodendro} also fits an ellipse for each individual structure and provides the major and minor axes of the fitted ellipse along with the position angle (PA) of its major axis. These parameters are important measures for the characterisation of the identified star-forming clumps.

\begin{figure}
\begin{center}
\includegraphics[width=3.2in]{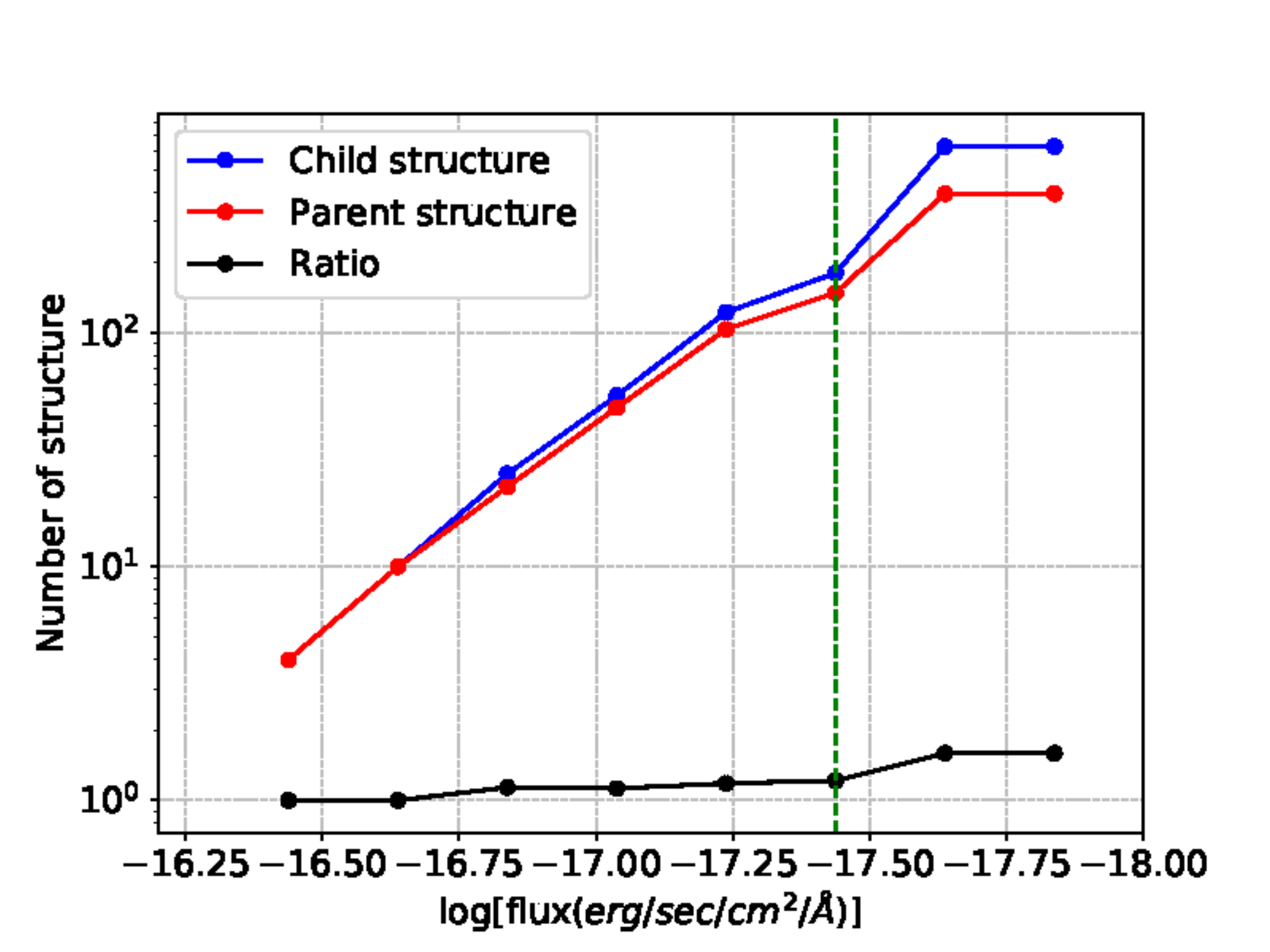} 
\includegraphics[width=3.2in]{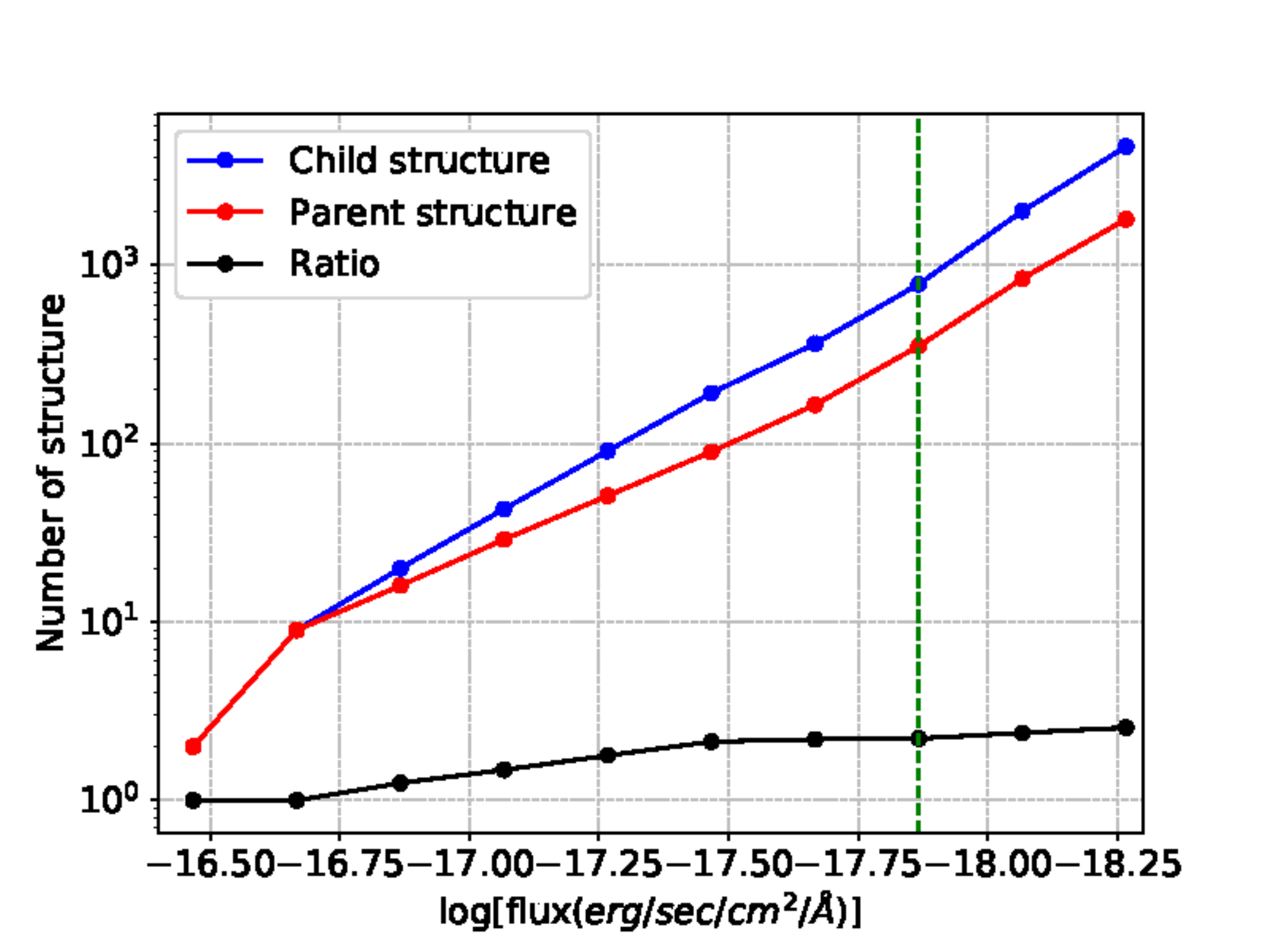}
 \caption{Number of child and parent structures identified in WLM (top panel) and IC~2574 (bottom panel) for varying threshold flux. The black lines show the ratio of child and parent structures. The vertical green dashed lines signify the value of threshold flux selected for each galaxy.}
 \label{tree_leaves}
 \end{center}
 \end{figure}

We fixed the minimum number of pixel for clump-identification as 10. This is so chosen that the size of the smallest clump will be similar to the FWHM of the PSF (which is $\sim$ 1.4$^{\prime\prime} \sim$ 3.4 pixels). At the distance of the galaxy WLM and IC~2574, this limiting size corresponds to $\sim$ 7 and $\sim$ 26 pc, respectively. To fix the threshold flux, we first examined the average background in each galaxy. We found the average FUV background flux (log[flux($erg/sec/cm^2/\AA{}$)]) for WLM and IC~2574 is $\sim$ $-18.54$ and $-18.96$, respectively. For selecting the threshold flux, we started with a flux value of 5 times the average background for each galaxy and increased it with 0.2 intervals in the logarithmic scale up to the value at which we detect less than 10 child structures. We identified both parent and child structures for these varying threshold fluxes and plotted the numbers in Figure \ref{tree_leaves}. For both the galaxies, we noticed the obvious trend of detecting less number of structures with increasing threshold flux. But the overall behaviour is not exactly similar in both the galaxies.

\begin{figure}

\includegraphics[width=3.4in]{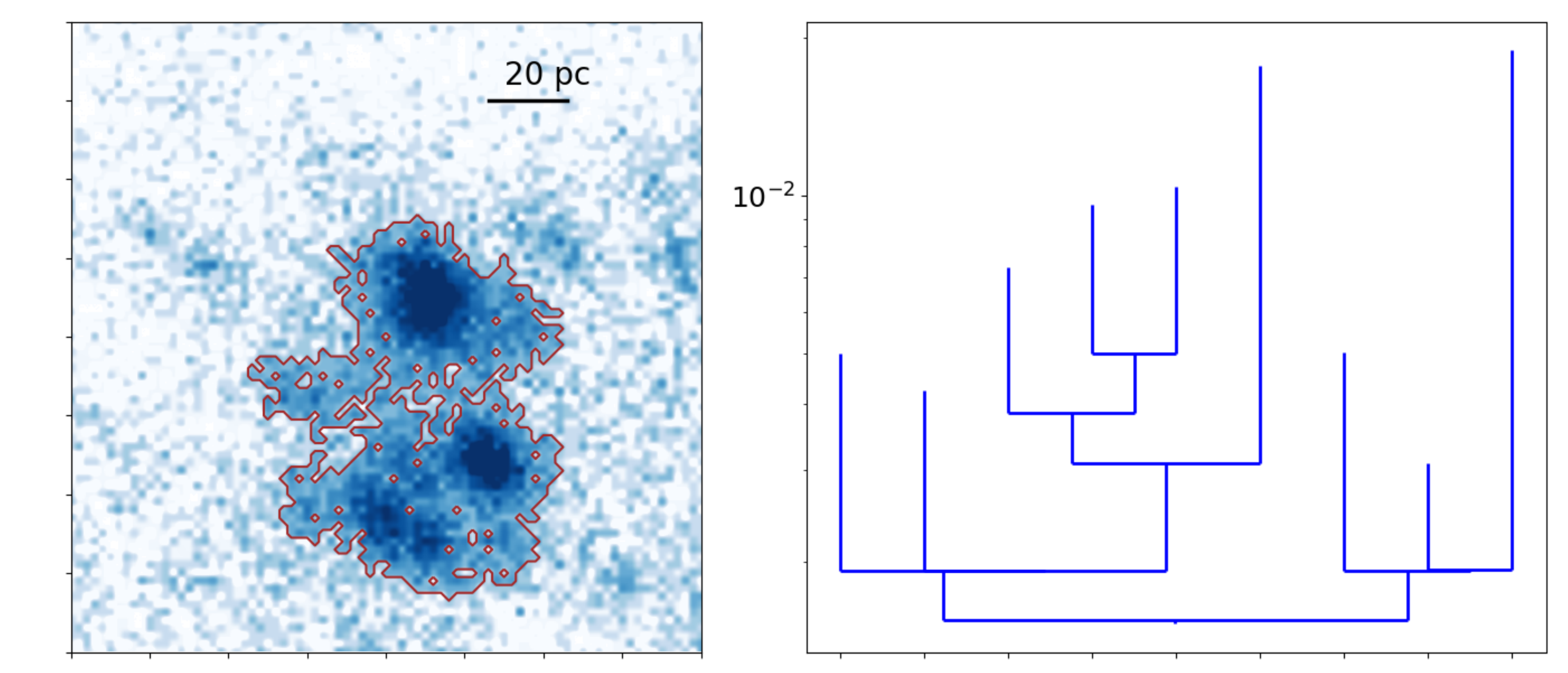} 
\includegraphics[width=3.4in]{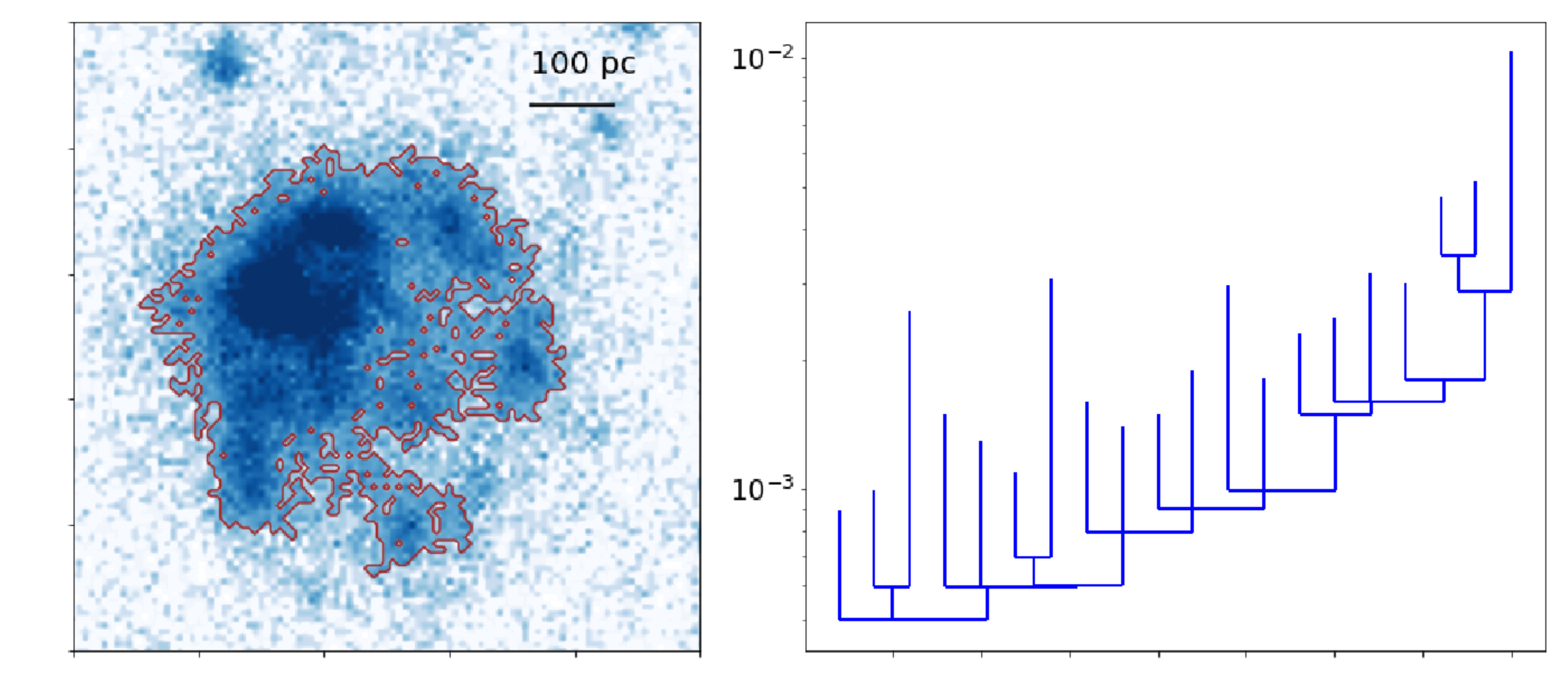} 
 \caption{Two selected star-forming regions (parent structure) from WLM (top panel) and IC~2574 (bottom panel) are shown along with their dendrograms which represent the overall structure tree. The brown contours show the boundary of the parent structures which host multiple child structure inside. Each branch of the dendrograms represents one child structure. The y-axis of the dendrogram shows the flux level in counts per second.}
 \label{dendrogram}
 \end{figure}

For the galaxy WLM, the number of parent and child structures identified for varying threshold flux is comparable (Figure \ref{tree_leaves} - top). The black line that shows the ratio of child to parent structures highlights the same. We noticed a slight jump in the black line at (log[flux($erg/$ $sec/cm^2/\AA{}$)] = $-17.44$). This flux value (which is $\sim$ 12 times the average background) is selected for studying the star-forming clumps in WLM. In the case of IC~2574 (Figure \ref{tree_leaves} - bottom), we noticed that with decreasing threshold flux, the ratio of child to parent structure gradually increases initially and then becomes relatively flat and further shows a small jump at log[flux($erg/$ $sec/cm^2/\AA{}$)] = $-17.87$ ($\sim$ 12 times the background flux). We selected this as the threshold flux for studying the star-forming clumps in IC~2574. A comparison between both the plots in Figure \ref{tree_leaves} signifies that the hierarchical splitting of star-forming regions for varying flux levels is more prominent in the galaxy IC 2574 than WLM. This may be due to the fact that IC 2574 is a bigger galaxy, leading to the formation of bigger structures, which then fragments into smaller clumps, whereas there are only moderately big structures in WLM. To highlight this more, we have selected one parent structure (among the brighter and larger ones) from each of the galaxies and shown them in Figure \ref{dendrogram} along with their dendrograms. The structure trees show the nature of hierarchical splitting from higher to lower flux levels above the selected thresholds in each galaxy. We noticed more substructures at different flux levels for the region in IC~2574 than WLM. 

\subsection{Properties of the clumps}
\label{s_property}
As our primary aim is to probe the nature of star-forming clumps up to smaller scales, we considered only the child structures, that are individual clump identified in the UVIT images for our further analysis. The number of child structures identified for the selected threshold flux in WLM and IC~2574 is 180 and 782, respectively. The {\it astrodendro} package provides several physical quantities for the identified structures. We used those parameters to estimate the size, ellipticity, and position angle of the identified clumps. Using the position, orientation, and size of the clumps, we also performed custom aperture photometry with {\it photutils} python package to estimate the flux of these clumps.\\

\textbf{Size}

\begin{figure}
\begin{center}
\includegraphics[width=3.2in]{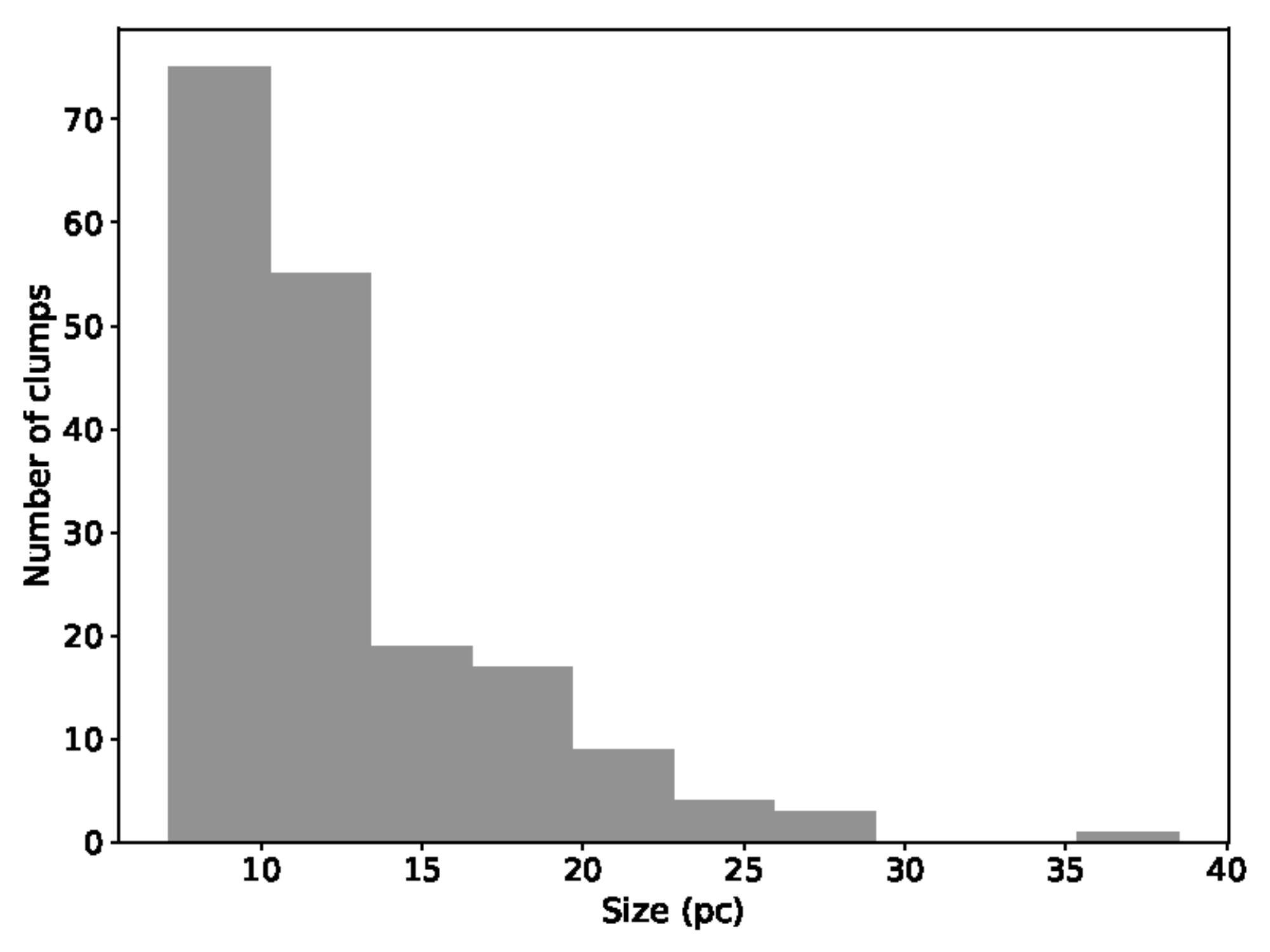} 
\includegraphics[width=3.2in]{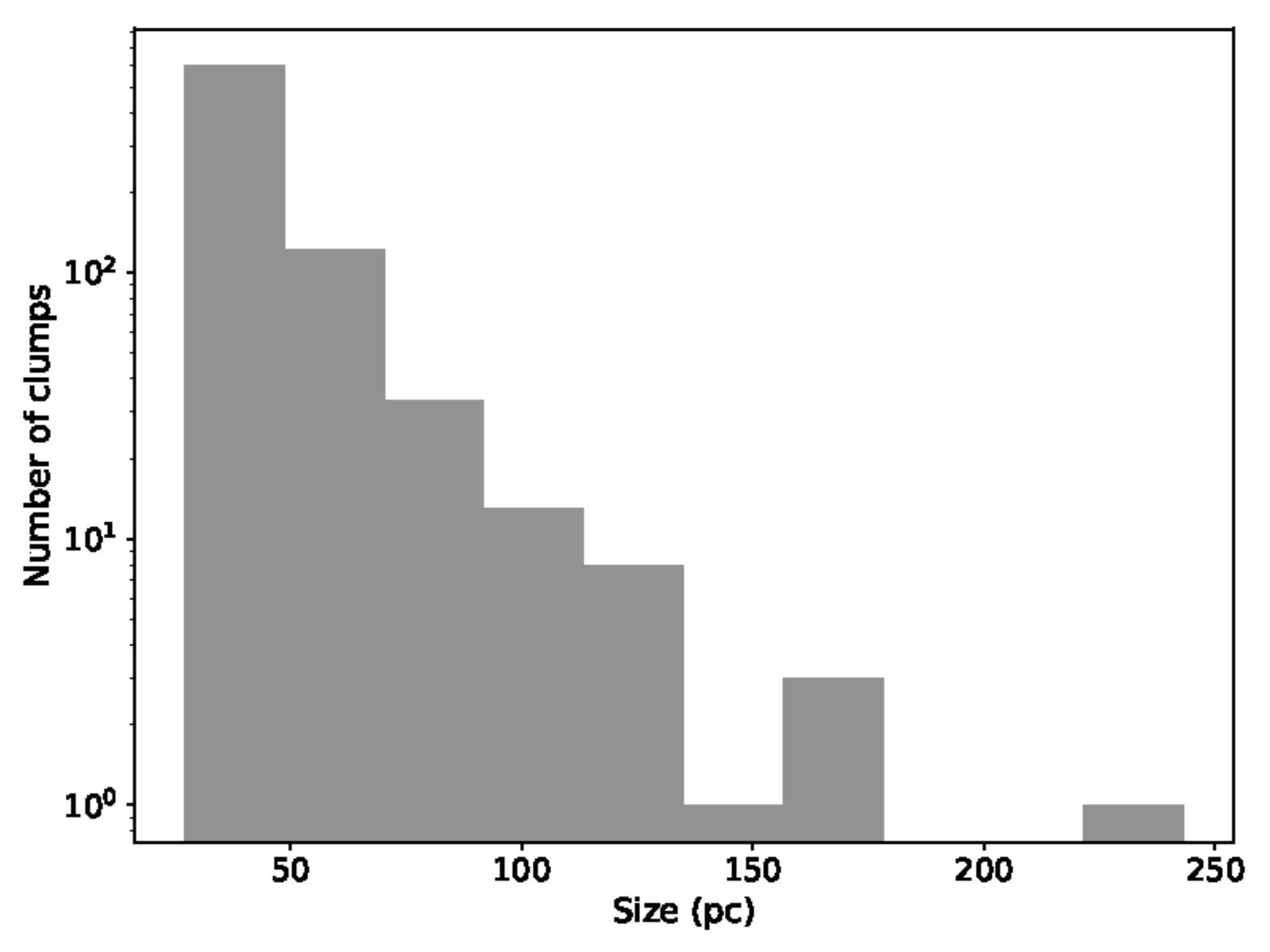}
 \caption{The histogram for size (in pc) of identified star-forming clumps. Top panel - WLM, bottom panel - IC~2574.}
 \label{clump_size}
 \end{center}
 \end{figure}

The clumps identified with {\it astrodendro} are mostly irregular in shape. We have shown a specific region of the galaxy WLM to highlight this in Figure \ref{wlm_clump_dendro}. To estimate the size of these clumps, we considered the area of each structure, provided by {\it astrodendro}, and equated it with the area of a circle with diameter d. The derived value of d is considered as the size of that clump. In Figure \ref{clump_size}, we have shown the histogram of the clump-size for both the galaxies. The FUV-bright clumps identified in WLM found to have sizes mostly in the range $\sim$ 7 - 30 pc, whereas IC~2574 hosts clumps mostly between $\sim$ 26 - 150 pc. We detected five larger clumps of size between 150 - 250 pc in IC~2574. The lower limit of the clump-size at the distance of each galaxy is limited by the UVIT resolution. To compare the nature of the histograms, we estimated standard deviation ($\sigma$) for both the distributions and found it to be 5.1 pc and 20.9 pc respectively for WLM and IC~2574. This highlights that the clump-size distribution in IC~2574 is broader than that in WLM. The smaller clumps dominate in number when compared to larger clumps in both the galaxies. It is also noticed that the galaxy WLM has produced relatively smaller clumps compared to IC~2574. Though, it is possible that the larger clumps identified in IC~2574, located around 4 times farther than WLM, are actually a combination of smaller clumps that could not be resolved further by UVIT.

As the two galaxies are located at different distances, the difference found in the overall range of estimated clump-size can be biased. To verify this, we degraded the resolution of WLM by placing it at the distance of IC~2574. We carried out the same analysis with {\it astrodendro} and identified clumps in the degraded WLM image. The detected clumps have a size mostly in the range 26 - 60 pc. Only a few clumps have a size between 60 - 90 pc. This overall picture is not a match to what we noticed in IC~2574. This signifies there is an intrinsic difference in the overall distribution of clump-size of the two galaxies and hence the difference found in the measured range is not purely a distance effect.\\

\textbf{Ellipticity}

\begin{figure*}
\begin{center}
\includegraphics[width=5.75in]{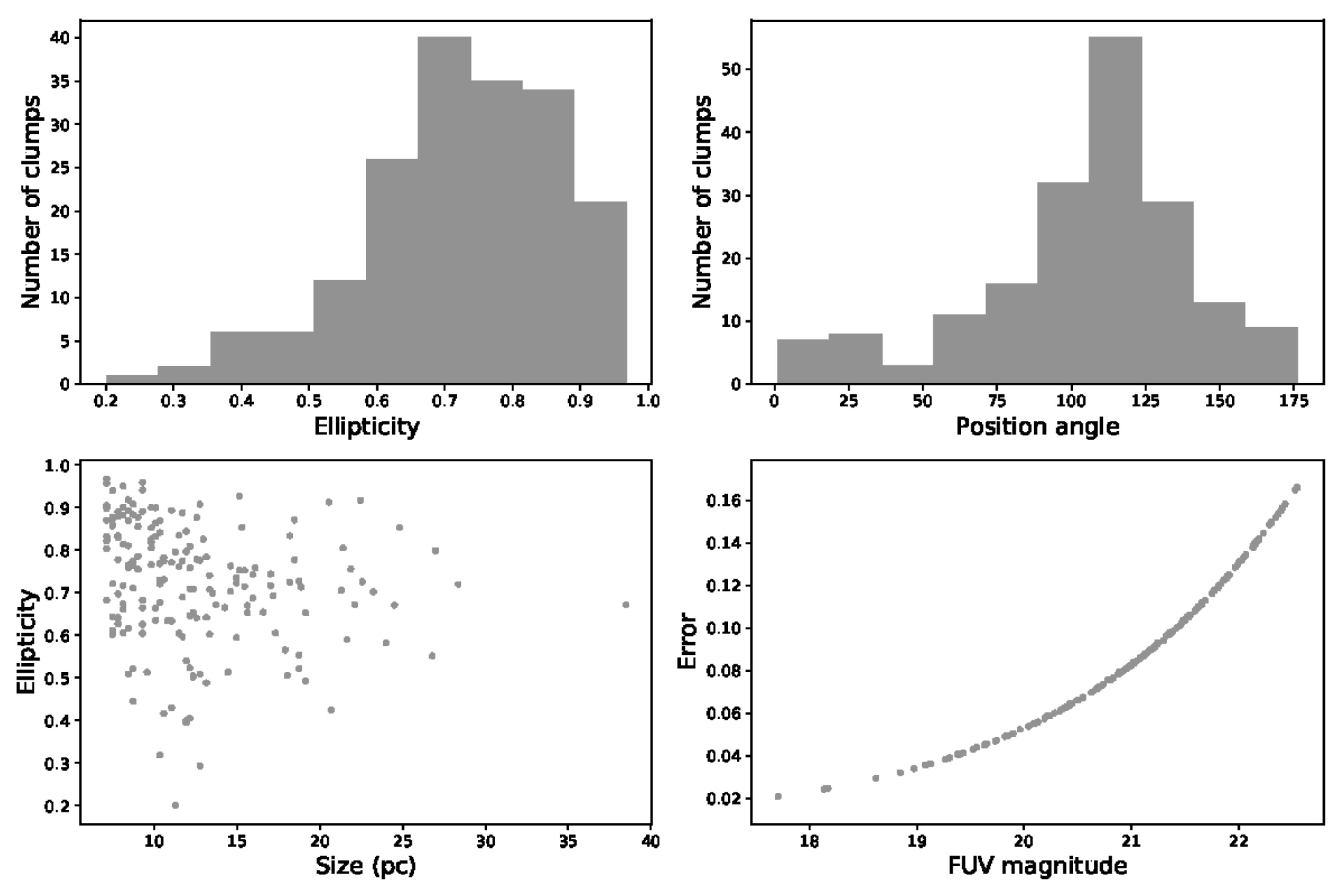} 
\caption{The properties of the star-forming clumps identified in the galaxy WLM. Top-left: Histogram of the ellipticity of identified clumps, Top-right: Histogram of the position angle of identified clumps, Bottom-left: The size and the ellipticity of the clumps, Bottom-right: Observed FUV magnitude of the clumps and the corresponding error.}
 \label{wlm_clumps}
 \end{center}
 \end{figure*}

\begin{figure*}
\begin{center}
\includegraphics[width=5.75in]{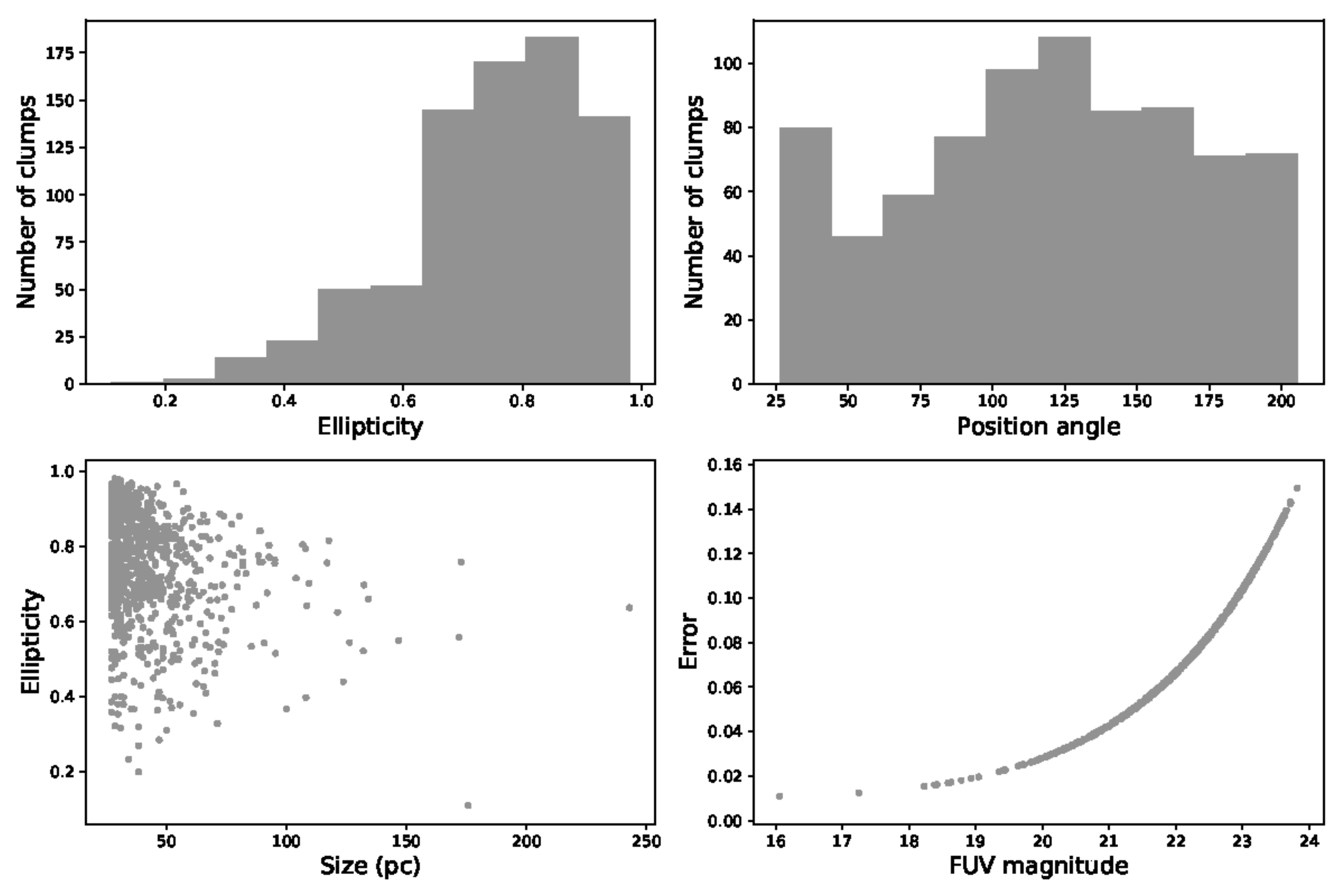} 
\caption{The properties of the star-forming clumps identified in the galaxy IC~2574. Each figure denotes the same as mentioned in Figure \ref{wlm_clumps}.}
 \label{ic2574_clumps}
 \end{center}
 \end{figure*}

We have estimated the ellipticity of the identified clumps (Figure \ref{wlm_clumps} and \ref{ic2574_clumps}). The {\it astrodendro} fits an ellipse inside the irregular-shaped boundary of each clump in the position-intensity plane and provides the length of major (a) and minor (b) axes of the fitted ellipse along with the position angle in the observed reference frame. We used the a, b values to estimate ellipticity of the clumps using $\epsilon = \sqrt{1 - \frac{b^2}{a^2}}$. A larger value of $\epsilon$ will signify a more elongated shape of the clump, whereas a smaller value will denote a more circular shape. In the top-left panel of Figure \ref{wlm_clumps} and \ref{ic2574_clumps}, we have shown the distribution of measured ellipticity for the clumps detected in WLM and IC~2574, respectively. We noticed that the majority of the star-forming clumps identified in both the galaxies are elongated in shape. The number of near-circular clumps is comparatively less in both the galaxies. The histogram for the galaxy WLM peaks at $\epsilon$ $\sim$ 0.7, whereas for IC~2574 the ellipticity value peaks at $\epsilon \sim$ 0.8.

We have shown the size and ellipticity of the identified clumps of both galaxies in \ref{wlm_clumps} and \ref{ic2574_clumps} (bottom-left). For both WLM and IC~2574, we noticed that most elliptic as well as circular structures are smaller in size. The larger clumps identified in WLM and IC~2574 found to have moderate ellipticity between $\sim$ 0.5 - 0.8.\\

\textbf{Position angle}

The estimated PA of the clumps are shown in top-right panel of Figure \ref{wlm_clumps} and \ref{ic2574_clumps}. The {\it astrodendro} measures PA with respect to the increasing x-axis in the observed frame. We converted these to the WCS frame and estimated position angle with respect to the west direction in a counter clock-wise direction. For example, a position angle in the range 0 $<$ PA $<$ 90 will mean the major axis of the clump is oriented along south-east to north-west direction, whereas a value between 90 $<$ PA $<$ 180 will signify the clump to be oriented along south-west to north-east. The results show that the majority of the clumps detected in the galaxy WLM have a specific mode in their orientation. This is clear from the distinct peak (around PA $\sim$ 110) noticed in the distribution of PA in Figure \ref{wlm_clumps} (top-right). On the other hand, the distribution is more uniform for the galaxy IC~2574 with a moderate peak around PA $\sim$ 125.\\

\textbf{Flux}

In order to characterise the star-forming clumps in terms of the FUV flux, we have performed custom aperture photometry using {\it photutils} package. We considered the position, major and minor axes (a,b), and the position angle of the fitted ellipse for each clump from {\it astrodendro} output. To measure the flux of a clump, we put an elliptic aperture on the clump center with major axis length as 2a and minor axis length as 2b with the same position angle. This is done to cover the identified irregular structure of each clump entirely. The measured flux is further corrected for the background and then converted to magnitude using the zero-points given in Table \ref{uvit_obs}. The apparent magnitudes of the identified clumps for galaxies WLM and IC~2574 are shown in the bottom-right of Figure \ref{wlm_clumps} and \ref{ic2574_clumps}, respectively. The clumps identified in WLM have a magnitude range between $\sim$ 18 - 22.5 mag, whereas in IC~2574 it ranges from $\sim$ 16 - 24 mag. We have also shown error for the measured magnitudes in the same figure.

\subsection{Radial distribution}
\begin{figure}
\begin{center}
\includegraphics[width=3.6in]{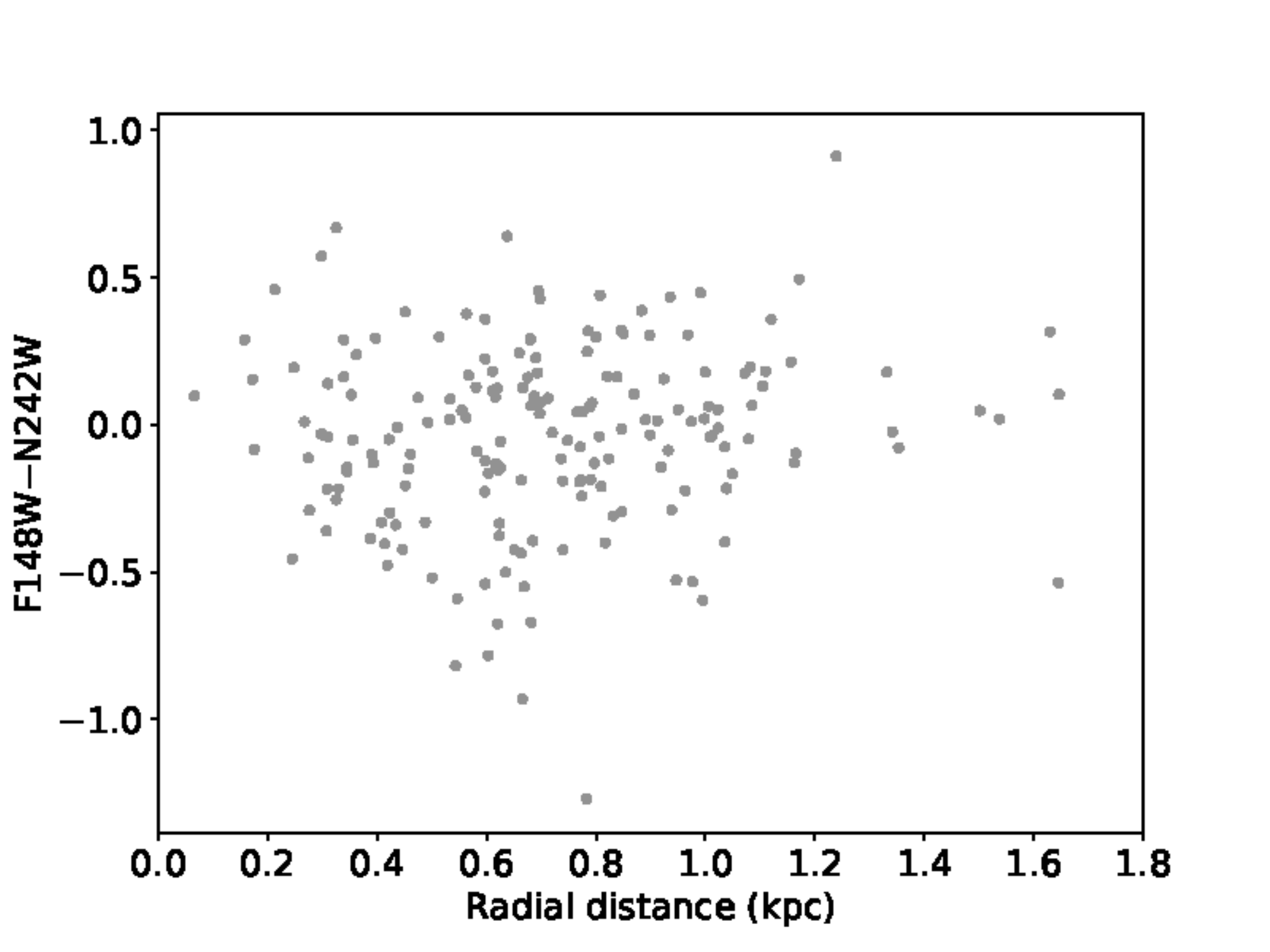} 
 \caption{The (F148W$-$N242W) colour and the galactocentric distance of the clumps identified in the galaxy WLM.}
 \label{radial}
 \end{center}
 \end{figure}

\begin{figure}
\begin{center}
\includegraphics[width=3.6in]{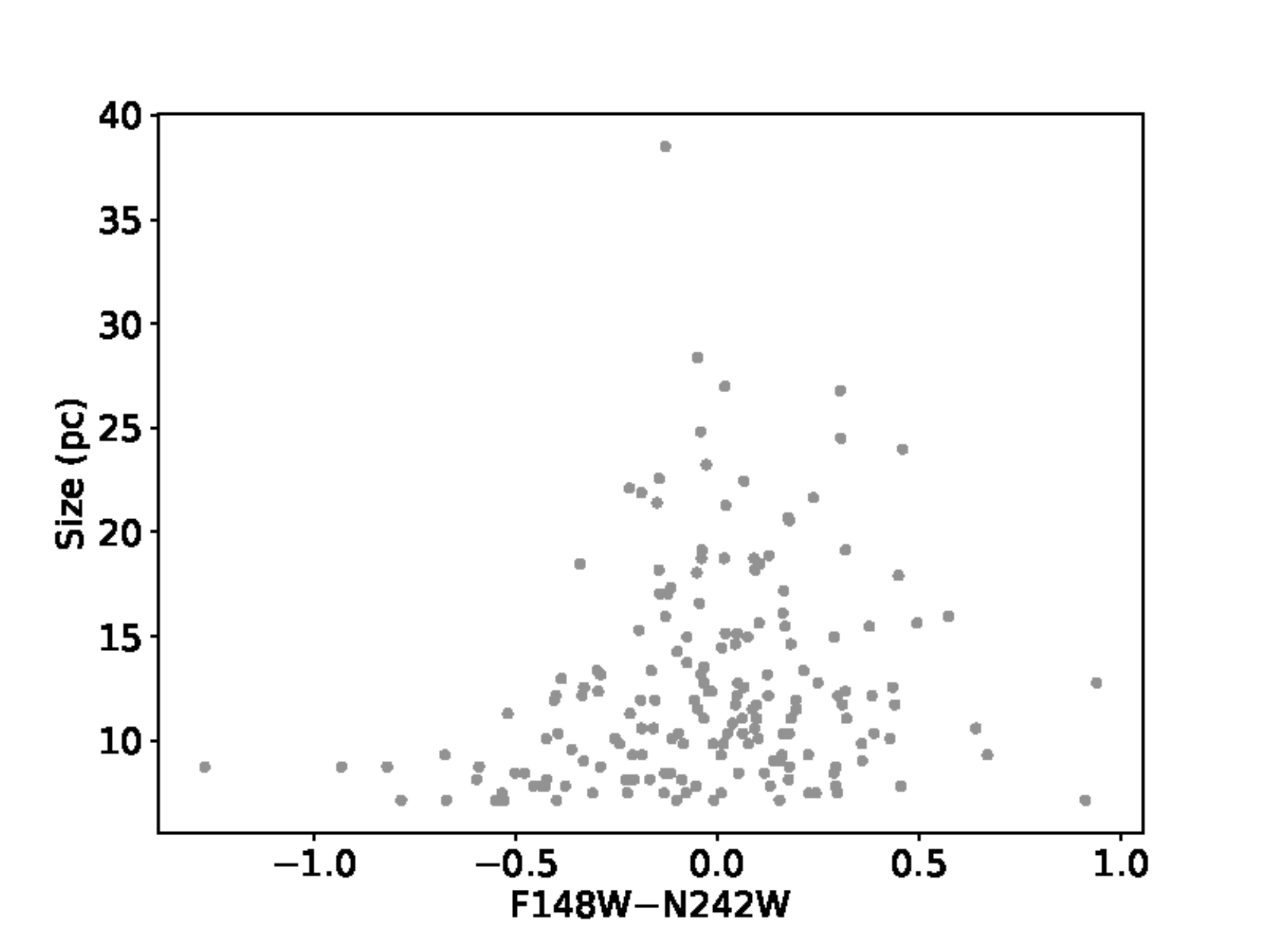} 
 \caption{The (F148W$-$N242W) colour and the size of the clumps identified in WLM.}
 \label{wlm_size_color}
 \end{center}
 \end{figure}

The (FUV$-$NUV) colour of a young star-forming clump is often used to trace the age of the stellar populations it contains. Simulation from the simple stellar population (SSP) models suggests that the UV colour becomes redder with increasing age of the SSP, and it is most sensitive up to a few hundred Myr \citep{goddard2010,mondal2019jaa}. The galaxy WLM has been observed in both F148W and N242W UVIT filters. In Section \ref{s_property}, we have estimated the FUV magnitudes of the identified star-forming clumps. We implemented the same positions and apertures, as derived from the F148W band image, on the N242W band image and estimated the NUV magnitudes of the clumps. The NUV magnitudes are also corrected for background emission. Using these magnitudes, we calculated (F148W$-$N242W) colour for each clump. We have also estimated the galactocentric distance to each identified clump following the steps discussed in \citet{mondal2019}. In Figure \ref{radial}, we have shown the (F148W$-$N242W) colour and the galactocentric distance (in kpc) of the clumps identified in the galaxy WLM. The clumps with bluer colour are seen between radii 0.4 - 0.8 kpc. These clumps are mostly located in the regions R1, R2, and R3, as shown in \citet{mondal2018}. The clumps identified in the central and the outer part of the galaxy are relatively redder in colour. We have also shown the (F148W$-$N242W) colour and the size of the detected clumps in Figure \ref{wlm_size_color}. We found that all the bluer clumps with (F148W$-$N242W) $< - 0.5$ are smaller in size ($<10$ pc). We also found a few redder clumps with (F148W$-$N242W) $> 0.5$) to be smaller than 10 pc in size. We note here that the photometric error in the (F148W$-$N242W) colour for clumps with extreme blue and red colours (which are smaller in size) is slightly higher (ranges between 0.2 - 0.4 mag). We could not study the (FUV$-$NUV) colour of the clumps in IC~2574, as the UVIT NUV data of the galaxy was not there.

\section{Results and Discussions}
\label{s_results}

The key aim of this study is to identify young star-forming regions in two nearby dwarf irregular galaxies and understand their properties. Despite having the same morphological class, the galaxy WLM and IC~2574 have significantly different characteristics. IC~2574 is around $\sim$ 5 times larger and $\sim$ 30 times massive than WLM. Though both the galaxies have low metallicity, IC~2574 is relatively metal-rich than WLM. Such contrast in the physical characteristics has motivated us to study the properties of star-forming clumps in each system and further compare them.

To identify star-forming clumps, we employed {\it astrodendro} python package. We varied the threshold flux in each galaxy and counted the number of child and parent structures for each different flux value. For the lower values of threshold flux, we found that the number of identified child structures is around 2 times higher than the parent structures in IC~2574, whereas for WLM it is around 1.5 times. This signifies the galaxy IC~2574 has more sub-structures at the lower flux levels compared to WLM. In other words, it suggests that the young star-forming regions in IC~2574 are more clumpy in nature. On the other hand, this could also be due to the observational bias as IC~2574 was imaged with 4 times deeper exposure than WLM and hence could detect many fainter clumps. We also note here that due to the deeper exposure, we expect to have less artefact in the identified clumps in IC~2574 than in WLM. With the selected threshold flux (i.e., $\sim$ 12 times the average background in each galaxy), we identified around 4 times more number of clumps in IC~2574 than WLM. This is mostly because of the larger size of the galaxy IC~2574. The difference in the exposure times can also be an add-on factor for such a contrast in the identified number. We also performed the same identification algorithm with the same thresholds and minimum pixel number on the galaxy images subtracted for smooth galaxy background (produced using {\it sep} python package \citep{bertin1996,barbary2016}) and identified 164 and 514 clumps, respectively in WLM and IC~2574. We searched for the same clumps as detected earlier and found that they have almost the same sizes with slightly less flux value ( which is mostly less than $\sim$ 20\% of the earlier value).

The identified clumps have sizes mostly between $\sim$ 7 - 30 pc in WLM and $\sim$ 26 - 150 pc in IC~2574. We could see that the bigger galaxy IC~2574 has clumps of larger size. The lower limit of the clump-size in each galaxy is fixed by the UVIT PSF and the distance to that galaxy. The overall size-range of the detected clumps is also sensitive to the distance of the galaxy \citep{bastian2007,garcia2010}. To confirm that the difference between the range of detected clump-size for two galaxies is not a result of the difference in their distances, we performed the same analysis on a smoothed image of WLM by pushing it at the distance of IC~2574. With this, we found the identified clumps to be smaller than 60 pc with a few between 60 - 90 pc. Therefore, the difference in the clump-size of two galaxies is mostly an intrinsic property and not an artefact. As FUV emission directly traces massive stellar populations, it is more likely that the star-forming clumps we identified in both the galaxies are OB associations. Several observations have targeted the OB associations in the Milky Way as well as in nearby spiral and dwarf galaxies and reported an average size between 15 - 100 pc \citep{melnik1995,ivanov1996,bresolin1996,bresolin1998,bastian2007,garcia2010}. The overall range of clump-size found in our study supports the earlier findings. \citet{pellerin2012} used HST optical observations for a limited region in the northern part of IC~2574 and detected 75 young stellar groups (age $\sim$ 10 Myr) with sizes between 10 - 120 pc, which matches well with our results. Although, it is possible that the larger clumps we detected are actually a combination of multiple smaller clumps that are not resolved by UVIT. The detected FUV clumps in WLM also show a good spatial correlation with the FUV bright stars as identified by HST \citep{bianchi2012}. In both the galaxies, we found the smaller clumps to dominate in numbers, which also have been reported earlier \citep{garcia2010}.

As the identified star-forming clumps have an irregular shape, we used {\it astrodendro} to fit ellipse for each of the structures and estimated their ellipticity. From the measured ellipticity values, we found the majority of the clumps to be elongated and very less have near-circular shape. This highlights the structure of young star-forming regions in these galaxies. The stellar groups identified by \citet{pellerin2012} in IC~2574 were found to have $\epsilon$ in the range 0 - 0.95. One possible reason for the elongated shape can be imaging in the FUV band. FUV emission mostly picks massive stars, which may not be distributed uniformly in a stellar group. Therefore, the shape of the clumps retrieved from the FUV imaging may not be the actual shape of the stellar group we are detecting. This also may be the reason why we see less number of near-circular clumps. The other important point to note here is the effect of galaxy inclination. Both the galaxies have relatively higher inclination angle (Table \ref{wlm_table}, \ref{ic2574_table}). For intrinsically non-spherical clumps, the inclination of the galaxy will have a clear effect on their shape, as seen in the sky plane.

Dwarf galaxies are systems that lack strong shear force and the presence of density waves. Such ordered force has a significant impact on the properties of molecular clouds and hence the star-forming clumps \citep{vogel1988,miyamoto2014}. The shape and orientation of the clumps can also be influenced by such organised force \citep{pellerin2012}. Here, we estimated the major axis PA of identified star-forming clumps to see whether there exists any specific trend in their orientation. Our analysis shows that the clumps detected in IC~2574 do not show any distinct mode in their orientation, whereas in WLM we noticed the majority of the clumps are aligned along south-west to north-east direction. Earlier studies by \citet{kepley2007,mondal2018} have shown the possibility of propagating star formation in the hook-like H~I structure around the centre in WLM. The specific mode of orientation of the star-forming clumps may have a connection with this. It may also be possible that the elongation is arising due to an underlying magnetic field or a filamentary structure in the molecular clouds.

The youngest clumps (bluer than $-$0.5 mag) in WLM are identified between radii 0.4 kpc to 0.8 kpc. As WLM has more star formation in the southern part, these clumps are located in the southern half of the galaxy, specifically in regions R1, R2, and R3 as defined in \citet{mondal2018}. Both the central and the outer parts of the galaxy host lesser clumps and they are mostly redder in colour. We also found that all the younger clumps with (F148W$-$N242W) $< -$0.5 are smaller than 10 pc in size. A few redder clumps of size $<$10 pc are also detected. We note here that the smaller clumps have a relatively larger photometric error in their measured magnitude. Therefore, the clumps with extreme colours will have a larger error ($\sim 0.2 - 0.4$) in their colour. We are unable to carry out a similar exercise for IC~2574 as NUV data is not available.

\section{Summary}
\label{s_summary}
The key results of the study are summarized below.

\begin{enumerate}
    \item We studied the characteristics of FUV-bright star-forming clumps in two nearby dwarf irregular galaxies WLM and IC~2574 using UVIT imaging observations.
    \item The identified clumps have a size between $\sim$ 7 - 30 pc in WLM and $\sim$ 26 - 150 pc in IC~2574. The average size of the clumps is larger in the galaxy IC~2574, which is bigger and massive than WLM.
    \item We found that the hierarchical splitting of star-forming regions is more prominent in IC~2574 than WLM.
    \item The young star-forming clumps identified in both the galaxies are mostly observed to be elongated in shape. 
    \item We did not find any specific orientation of the clump major axis in IC~2574, whereas in WLM the majority of the clumps are oriented along south-west to north-east direction.
    \item The youngest star-forming clumps in WLM are detected between radii 0.4 kpc to 0.8 kpc. Both the central and outer parts of the galaxy are relatively less active in recent times.
    
\end{enumerate}

\section*{Acknowledgements}

This publication uses the data from the AstroSat mission of the Indian Space Research Organisation (ISRO), archived at the Indian Space Science Data Centre (ISSDC). This publication uses UVIT data processed by the payload operations centre at IIA. The UVIT is built in collaboration between IIA, IUCAA, TIFR, ISRO, and CSA. Indian Institutions and the Canadian Space Agency have contributed to the work presented in this paper. Several groups from ISAC (ISRO), Bengaluru, and IISU (ISRO), Trivandrum have contributed to the design, fabrication, and testing of the payload. The Mission Group (ISAC) and
ISTRAC (ISAC) continue to provide support in making observations with, and reception and initial processing of the data. We gratefully thank all the individuals involved in the various teams for providing their support to the project from the early stages of the design to launch and observations with it in the orbit. This research made use of Matplotlib (Hunter 2007), Astropy (Astropy Collaboration et al. 2013, 2018), Astrodendro (http://www.dendrograms.org/), community-developed core Python packages for Astronomy. Finally, we thank the referees for the valuable suggestions.


\vspace{-1em}



\begin{thebibliography}{}
\expandafter\ifx\csname natexlab\endcsname\relax\def\natexlab#1{#1}\fi
\providecommand{\url}[1]{\href{#1}{#1}}

\bibitem[{{Astropy Collaboration} {et~al.}(2013){Astropy Collaboration},
  {Robitaille}, {Tollerud}, {Greenfield}, {Droettboom}, {Bray}, {Aldcroft},
  {Davis}, {Ginsburg}, {Price-Whelan}, {Kerzendorf}, {Conley}, {Crighton},
  {Barbary}, {Muna}, {Ferguson}, {Grollier}, {Parikh}, {Nair}, {Unther},
  {Deil}, {Woillez}, {Conseil}, {Kramer}, {Turner}, {Singer}, {Fox}, {Weaver},
  {Zabalza}, {Edwards}, {Azalee Bostroem}, {Burke}, {Casey}, {Crawford},
  {Dencheva}, {Ely}, {Jenness}, {Labrie}, {Lim}, {Pierfederici}, {Pontzen},
  {Ptak}, {Refsdal}, {Servillat}, \& {Streicher}}]{astropy2013}
{Astropy Collaboration}, {Robitaille}, T.~P., {Tollerud}, E.~J., {et~al.} 2013,
  A\&A, 558, A33

\bibitem[{{Astropy Collaboration} {et~al.}(2018){Astropy Collaboration},
  {Price-Whelan}, {Sip{\H o}cz}, {G{\"u}nther}, {Lim}, {Crawford}, {Conseil},
  {Shupe}, {Craig}, {Dencheva}, {Ginsburg}, {VanderPlas}, {Bradley},
  {P{\'e}rez-Su{\'a}rez}, {de Val-Borro}, {Aldcroft}, {Cruz}, {Robitaille},
  {Tollerud}, {Ardelean}, {Babej}, {Bach}, {Bachetti}, {Bakanov}, {Bamford},
  {Barentsen}, {Barmby}, {Baumbach}, {Berry}, {Biscani}, {Boquien}, {Bostroem},
  {Bouma}, {Brammer}, {Bray}, {Breytenbach}, {Buddelmeijer}, {Burke},
  {Calderone}, {Cano Rodr{\'{\i}}guez}, {Cara}, {Cardoso}, {Cheedella},
  {Copin}, {Corrales}, {Crichton}, {D'Avella}, {Deil}, {Depagne}, {Dietrich},
  {Donath}, {Droettboom}, {Earl}, {Erben}, {Fabbro}, {Ferreira}, {Finethy},
  {Fox}, {Garrison}, {Gibbons}, {Goldstein}, {Gommers}, {Greco}, {Greenfield},
  {Groener}, {Grollier}, {Hagen}, {Hirst}, {Homeier}, {Horton}, {Hosseinzadeh},
  {Hu}, {Hunkeler}, {Ivezi{\'c}}, {Jain}, {Jenness}, {Kanarek}, {Kendrew},
  {Kern}, {Kerzendorf}, {Khvalko}, {King}, {Kirkby}, {Kulkarni}, {Kumar},
  {Lee}, {Lenz}, {Littlefair}, {Ma}, {Macleod}, {Mastropietro}, {McCully},
  {Montagnac}, {Morris}, {Mueller}, {Mumford}, {Muna}, {Murphy}, {Nelson},
  {Nguyen}, {Ninan}, {N{\"o}the}, {Ogaz}, {Oh}, {Parejko}, {Parley}, {Pascual},
  {Patil}, {Patil}, {Plunkett}, {Prochaska}, {Rastogi}, {Reddy Janga},
  {Sabater}, {Sakurikar}, {Seifert}, {Sherbert}, {Sherwood-Taylor}, {Shih},
  {Sick}, {Silbiger}, {Singanamalla}, {Singer}, {Sladen}, {Sooley},
  {Sornarajah}, {Streicher}, {Teuben}, {Thomas}, {Tremblay}, {Turner},
  {Terr{\'o}n}, {van Kerkwijk}, {de la Vega}, {Watkins}, {Weaver}, {Whitmore},
  {Woillez}, {Zabalza}, \& {Astropy Contributors}}]{astropy2018}
{Astropy Collaboration}, {Price-Whelan}, A.~M., {Sip{\H o}cz}, B.~M., {et~al.}
  2018, AJ, 156, 123

\bibitem[{Barbary(2016)}]{barbary2016}
Barbary, K. 2016, Journal of Open Source Software, 1, 58

\bibitem[{{Bastian} {$et~al$.}(2007){Bastian}, {Ercolano}, {Gieles},
  {Rosolowsky}, {Scheepmaker}, {Gutermuth}, \& {Efremov}}]{bastian2007}
{Bastian}, N., {Ercolano}, B., {Gieles}, M., {$et~al$.} 2007, MNRAS, 379, 1302

\bibitem[{{Bertin} \& {Arnouts}(1996)}]{bertin1996}
{Bertin}, E., \& {Arnouts}, S. 1996, A\&AS, 117, 393

\bibitem[{{Bianchi} {et~al.}(2012{\natexlab{a}}){Bianchi}, {Efremova}, {Hodge},
  \& {Kang}}]{bianchi2012a}
{Bianchi}, L., {Efremova}, B., {Hodge}, P., \& {Kang}, Y. 2012{\natexlab{a}},
  AJ, 144, 142

\bibitem[{{Bianchi} {et~al.}(2012{\natexlab{b}}){Bianchi}, {Efremova}, {Hodge},
  {Massey}, \& {Olsen}}]{bianchi2012}
{Bianchi}, L., {Efremova}, B., {Hodge}, P., {Massey}, P., \& {Olsen}, K.~A.~G.
  2012{\natexlab{b}}, AJ, 143, 74

\bibitem[{{Bianchi} {et~al.}(2014){Bianchi}, {Kang}, {Hodge}, {Dalcanton}, \&
  {Williams}}]{bianchi2014c}
{Bianchi}, L., {Kang}, Y., {Hodge}, P., {Dalcanton}, J., \& {Williams}, B.
  2014, Advances in Space Research, 53, 928


\bibitem[{{Bresolin} {et~al.}(1996){Bresolin}, {Kennicutt}, \&
  {Stetson}}]{bresolin1996}
{Bresolin}, F., {Kennicutt}, Jr., R.~C., \& {Stetson}, P.~B. 1996, AJ, 112,
  1009

\bibitem[{{Bresolin} {et~al.}(1998){Bresolin}, {Kennicutt}, {Ferrarese},
  {Gibson}, {Graham}, {Macri}, {Phelps}, {Rawson}, {Sakai}, {Silbermann},
  {Stetson}, \& {Turner}}]{bresolin1998}
{Bresolin}, F., {Kennicutt}, Jr., R.~C., {Ferrarese}, L., {et~al.} 1998, AJ,
  116, 119

\bibitem[{{Calzetti} {et~al.}(2015){Calzetti}, {Lee}, {Sabbi}, {Adamo},
  {Smith}, {Andrews}, {Ubeda}, {Bright}, {Thilker}, {Aloisi}, {Brown},
  {Chandar}, {Christian}, {Cignoni}, {Clayton}, {da Silva}, {de Mink}, {Dobbs},
  {Elmegreen}, {Elmegreen}, {Evans}, {Fumagalli}, {Gallagher}, {Gouliermis},
  {Grebel}, {Herrero}, {Hunter}, {Johnson}, {Kennicutt}, {Kim}, {Krumholz},
  {Lennon}, {Levay}, {Martin}, {Nair}, {Nota}, {{\"O}stlin}, {Pellerin},
  {Prieto}, {Regan}, {Ryon}, {Schaerer}, {Schiminovich}, {Tosi}, {Van Dyk},
  {Walterbos}, {Whitmore}, \& {Wofford}}]{calzetti2015}
{Calzetti}, D., {Lee}, J.~C., {Sabbi}, E., {et~al.} 2015, AJ, 149, 51

\bibitem[{{Cannon} {et~al.}(2005){Cannon}, {Walter}, {Bendo}, {Calzetti},
  {Dale}, {Draine}, {Engelbracht}, {Gordon}, {Helou}, {Kennicutt}, {Murphy},
  {Thornley}, {Armus}, {Hollenbach}, {Leitherer}, {Regan}, {Roussel}, \&
  {Sheth}}]{cannon2005}
{Cannon}, J.~M., {Walter}, F., {Bendo}, G.~J., {et~al.} 2005, ApJL, 630, L37

\bibitem[{{Cignoni} {et~al.}(2018){Cignoni}, {Sacchi}, {Aloisi}, {Tosi},
  {Calzetti}, {Lee}, {Sabbi}, {Adamo}, {Cook}, {Dale}, {Elmegreen},
  {Gallagher}, {Gouliermis}, {Grasha}, {Grebel}, {Hunter}, {Johnson}, {Messa},
  {Smith}, {Thilker}, {Ubeda}, \& {Whitmore}}]{cignoni2018}
{Cignoni}, M., {Sacchi}, E., {Aloisi}, A., {et~al.} 2018, ApJ, 856, 62

\bibitem[{{Cignoni} {et~al.}(2019){Cignoni}, {Sacchi}, {Tosi}, {Aloisi},
  {Cook}, {Calzetti}, {Lee}, {Sabbi}, {Thilker}, {Adamo}, {Dale}, {Elmegreen},
  {Gallagher}, {Grebel}, {Johnson}, {Messa}, {Smith}, \& {Ubeda}}]{cignoni2019}
{Cignoni}, M., {Sacchi}, E., {Tosi}, M., {et~al.} 2019, ApJ, 887, 112

\bibitem[{{Dalcanton} {et~al.}(2009){Dalcanton}, {Williams}, {Seth}, {Dolphin},
  {Holtzman}, {Rosema}, {Skillman}, {Cole}, {Girardi}, {Gogarten},
  {Karachentsev}, {Olsen}, {Weisz}, {Christensen}, {Freeman}, {Gilbert},
  {Gallart}, {Harris}, {Hodge}, {de Jong}, {Karachentseva}, {Mateo}, {Stetson},
  {Tavarez}, {Zaritsky}, {Governato}, \& {Quinn}}]{dalcanton2009}
{Dalcanton}, J.~J., {Williams}, B.~F., {Seth}, A.~C., {et~al.} 2009, ApJS,
  183, 67

\bibitem[{{de Vaucouleurs} {et~al.}(1991){de Vaucouleurs}, {de Vaucouleurs},
  {Corwin}, {Buta}, {Paturel}, \& {Fouque}}]{devaucouleurs1991b}
{de Vaucouleurs}, G., {de Vaucouleurs}, A., {Corwin}, Herold~G., J., {et~al.}
  1991, {Third Reference Catalogue of Bright Galaxies}

\bibitem[{{Elmegreen} {et~al.}(2000){Elmegreen}, {Efremov}, {Pudritz}, \&
  {Zinnecker}}]{elmegreen2000}
{Elmegreen}, B.~G., {Efremov}, Y., {Pudritz}, R.~E., \& {Zinnecker}, H. 2000,
  Protostars and Planets IV, 179

\bibitem[{{Gallouet} {et~al.}(1975){Gallouet}, {Heidmann}, \&
  {Dampierre}}]{gallouet1975}
{Gallouet}, L., {Heidmann}, N., \& {Dampierre}, F. 1975, A\&AS, 19, 1

\bibitem[{{Garcia} {et~al.}(2010){Garcia}, {Herrero}, {Castro}, {Corral}, \&
  {Rosenberg}}]{garcia2010}
{Garcia}, M., {Herrero}, A., {Castro}, N., {Corral}, L., \& {Rosenberg}, A.
  2010, A\&A, 523, A23

\bibitem[{{Gerola} {et~al.}(1980){Gerola}, {Seiden}, \&
  {Schulman}}]{gerola1980}
{Gerola}, H., {Seiden}, P.~E., \& {Schulman}, L.~S. 1980, ApJ, 242, 517

\bibitem[{{Gil de Paz} {et~al.}(2007){Gil de Paz}, {Boissier}, {Madore},
  {Seibert}, {Joe}, {Boselli}, {Wyder}, {Thilker}, {Bianchi}, {Rey}, {Rich},
  {Barlow}, {Conrow}, {Forster}, {Friedman}, {Martin}, {Morrissey}, {Neff},
  {Schiminovich}, {Small}, {Donas}, {Heckman}, {Lee}, {Milliard}, {Szalay}, \&
  {Yi}}]{gildepaz2007}
{Gil de Paz}, A., {Boissier}, S., {Madore}, B.~F., {et~al.} 2007, ApJS, 173,
  185

\bibitem[{{Girish} {et~al.}(2017){Girish}, {Tandon}, {Sriram}, {Kumar}, \&
  {Postma}}]{girish2017}
{Girish}, V., {Tandon}, S.~N., {Sriram}, S., {Kumar}, A., \& {Postma}, J. 2017,
  Experimental Astronomy, 43, 59

\bibitem[{{Goddard} {et~al.}(2010){Goddard}, {Kennicutt}, \&
  {Ryan-Weber}}]{goddard2010}
{Goddard}, Q.~E., {Kennicutt}, R.~C., \& {Ryan-Weber}, E.~V. 2010, MNRAS, 405,
  2791

\bibitem[{{Grasha} {et~al.}(2017){Grasha}, {Elmegreen}, {Calzetti}, {Adamo},
  {Aloisi}, {Bright}, {Cook}, {Dale}, {Fumagalli}, {Gallagher}, {Gouliermis},
  {Grebel}, {Kahre}, {Kim}, {Krumholz}, {Lee}, {Messa}, {Ryon}, \&
  {Ubeda}}]{grasha2017}
{Grasha}, K., {Elmegreen}, B.~G., {Calzetti}, D., {et~al.} 2017, ApJ, 842, 25

\bibitem[{{Hunter}(1997)}]{hunter1997}
{Hunter}, D. 1997, PASP, 109, 937

\bibitem[{Hunter(2007)}]{matplotlib2007}
Hunter, J.~D. 2007, Computing In Science \& Engineering, 9, 90

\bibitem[{{Ivanov}(1996)}]{ivanov1996}
{Ivanov}, G.~R. 1996, A\&A, 305, 708

\bibitem[{{Kang} {et~al.}(2009){Kang}, {Bianchi}, \& {Rey}}]{kang2009}
{Kang}, Y., {Bianchi}, L., \& {Rey}, S.-C. 2009, ApJ, 703, 614

\bibitem[{{Kennicutt} \& {Evans}(2012)}]{kennicutt2012}
{Kennicutt}, R.~C., \& {Evans}, N.~J. 2012, ARA\&A, 50, 531

\bibitem[{{Kepley} {et~al.}(2007){Kepley}, {Wilcots}, {Hunter}, \&
  {Nordgren}}]{kepley2007}
{Kepley}, A.~A., {Wilcots}, E.~M., {Hunter}, D.~A., \& {Nordgren}, T. 2007,
  AJ, 133, 2242

\bibitem[{{Kumar} {et~al.}(2012){Kumar}, {Ghosh}, {Hutchings}, {Kamath},
  {Kathiravan}, {Mahesh}, {Murthy}, {Nagbhushana}, {Pati}, {Rao}, {Rao},
  {Sriram}, \& {Tandon}}]{kumar2012}
{Kumar}, A., {Ghosh}, S.~K., {Hutchings}, J., {et~al.} 2012, in proc SPIE, Vol.
  8443, Space Telescopes and Instrumentation 2012: Ultraviolet to Gamma Ray,
  84431N

\bibitem[{{Leaman} {et~al.}(2012){Leaman}, {Venn}, {Brooks}, {Battaglia},
  {Cole}, {Ibata}, {Irwin}, {McConnachie}, {Mendel}, \& {Tolstoy}}]{leaman2012}
{Leaman}, R., {Venn}, K.~A., {Brooks}, A.~M., {et~al.} 2012, ApJ, 750, 33

\bibitem[{{Lee} {et~al.}(2006){Lee}, {Skillman}, {Cannon}, {Jackson}, {Gehrz},
  {Polomski}, \& {Woodward}}]{lee2006}
{Lee}, H., {Skillman}, E.~D., {Cannon}, J.~M., {et~al.} 2006, ApJ, 647, 970

\bibitem[{{Martin} {et~al.}(2005){Martin}, {Fanson}, {Schiminovich},
  {Morrissey}, {Friedman}, {Barlow}, {Conrow}, {Grange}, {Jelinsky},
  {Milliard}, {Siegmund}, {Bianchi}, {Byun}, {Donas}, {Forster}, {Heckman},
  {Lee}, {Madore}, {Malina}, {Neff}, {Rich}, {Small}, {Surber}, {Szalay},
  {Welsh}, \& {Wyder}}]{martin2005}
{Martin}, D.~C., {Fanson}, J., {Schiminovich}, D., {et~al.} 2005, ApJL, 619,
  L1

\bibitem[{{McQuinn} {et~al.}(2015){McQuinn}, {Skillman}, {Dolphin}, \&
  {Mitchell}}]{mcquinn2015}
{McQuinn}, K.~B.~W., {Skillman}, E.~D., {Dolphin}, A.~E., \& {Mitchell}, N.~P.
  2015, ApJ, 808, 109

\bibitem[{{Melena} {et~al.}(2009){Melena}, {Elmegreen}, {Hunter}, \&
  {Zernow}}]{melena2009}
{Melena}, N.~W., {Elmegreen}, B.~G., {Hunter}, D.~A., \& {Zernow}, L. 2009,
  AJ, 138, 1203

\bibitem[{{Mel'Nik} \& {Efremov}(1995)}]{melnik1995}
{Mel'Nik}, A.~M., \& {Efremov}, Y.~N. 1995, Astronomy Letters, 21, 10

\bibitem[{{Miyamoto} {et~al.}(2014){Miyamoto}, {Nakai}, \&
  {Kuno}}]{miyamoto2014}
{Miyamoto}, Y., {Nakai}, N., \& {Kuno}, N. 2014, PASJ, 66, 36

\bibitem[{{Mondal} {et~al.}(2018){Mondal}, {Subramaniam}, \&
  {George}}]{mondal2018}
{Mondal}, C., {Subramaniam}, A., \& {George}, K. 2018, AJ, 156, 109

\bibitem[{{Mondal} {et~al.}(2019){Mondal}, {Subramaniam}, \&
  {George}}]{mondal2019}
{Mondal}, C., {Subramaniam}, A., \& {George}, K. 2019, AJ, 158, 229

\bibitem[{{Mondal} {et~al.}(2019){Mondal}, {Subramaniam}, \&
  {George}}]{mondal2019jaa}
{Mondal}, C., {Subramaniam}, A., \& {George}, K. 2019, Journal of Astrophysics and Astronomy, 40, 35

\bibitem[{{Pasquali} {et~al.}(2008){Pasquali}, {Leroy}, {Rix}, {Walter},
  {Herbst}, {Giallongo}, {Ragazzoni}, {Baruffolo}, {Speziali}, {Hill},
  {Beccari}, {Bouch{\'e}}, {Buschkamp}, {Kochanek}, {Skillman}, \&
  {Bechtold}}]{pasquali2008}
{Pasquali}, A., {Leroy}, A., {Rix}, H.-W., {et~al.} 2008, ApJ, 687, 1004

\bibitem[{{Patton} {et~al.}(2013){Patton}, {Torrey}, {Ellison}, {Mendel}, \&
  {Scudder}}]{patton2013}
{Patton}, D.~R., {Torrey}, P., {Ellison}, S.~L., {Mendel}, J.~T., \& {Scudder},
  J.~M. 2013, MNRAS, 433, L59

\bibitem[{{Pellerin} {et~al.}(2012){Pellerin}, {Meyer}, {Calzetti}, \&
  {Harris}}]{pellerin2012}
{Pellerin}, A., {Meyer}, M.~M., {Calzetti}, D., \& {Harris}, J. 2012, AJ, 144,
  182

\bibitem[{{Postma} {et~al.}(2011){Postma}, {Hutchings}, \&
  {Leahy}}]{postma2011}
{Postma}, J., {Hutchings}, J.~B., \& {Leahy}, D. 2011, PASP, 123, 833

\bibitem[{{Postma} \& {Leahy}(2017)}]{postma2017}
{Postma}, J.~E., \& {Leahy}, D. 2017, PASP, 129, 115002

\bibitem[{{Skrutskie} {et~al.}(2006){Skrutskie}, {Cutri}, {Stiening},
  {Weinberg}, {Schneider}, {Carpenter}, {Beichman}, {Capps}, {Chester},
  {Elias}, {Huchra}, {Liebert}, {Lonsdale}, {Monet}, {Price}, {Seitzer},
  {Jarrett}, {Kirkpatrick}, {Gizis}, {Howard}, {Evans}, {Fowler}, {Fullmer},
  {Hurt}, {Light}, {Kopan}, {Marsh}, {McCallon}, {Tam}, {Van Dyk}, \&
  {Wheelock}}]{skrutskie2006}
{Skrutskie}, M.~F., {Cutri}, R.~M., {Stiening}, R., {et~al.} 2006, AJ, 131,
  1163

\bibitem[{{Stierwalt} {et~al.}(2015){Stierwalt}, {Besla}, {Patton}, {Johnson},
  {Kallivayalil}, {Putman}, {Privon}, \& {Ross}}]{stierwalt2015}
{Stierwalt}, S., {Besla}, G., {Patton}, D., {et~al.} 2015, ApJ, 805, 2

\bibitem[{{Tandon} {et~al.}(2017){Tandon}, {Subramaniam}, {Girish}, {Postma},
  {Sankarasubramanian}, {Sriram}, {Stalin}, {Mondal}, {Sahu}, {Joseph},
  {Hutchings}, {Ghosh}, {Barve}, {George}, {Kamath}, {Kathiravan}, {Kumar},
  {Lancelot}, {Leahy}, {Mahesh}, {Mohan}, {Nagabhushana}, {Pati}, {Kameswara
  Rao}, {Sreedhar}, \& {Sreekumar}}]{tandon2017}
{Tandon}, S.~N., {Subramaniam}, A., {Girish}, V., {et~al.} 2017, AJ, 154, 128

\bibitem[{{Thilker} {et~al.}(2007){Thilker}, {Bianchi}, {Meurer}, {Gil de Paz},
  {Boissier}, {Madore}, {Boselli}, {Ferguson}, {Mu{\~n}oz-Mateos}, {Madsen},
  {Hameed}, {Overzier}, {Forster}, {Friedman}, {Martin}, {Morrissey}, {Neff},
  {Schiminovich}, {Seibert}, {Small}, {Wyder}, {Donas}, {Heckman}, {Lee},
  {Milliard}, {Rich}, {Szalay}, {Welsh}, \& {Yi}}]{thilker2007}
{Thilker}, D.~A., {Bianchi}, L., {Meurer}, G., {et~al.} 2007, ApJS, 173, 538

\bibitem[{{Tolstoy} {et~al.}(2009){Tolstoy}, {Hill}, \& {Tosi}}]{tolstoy2009}
{Tolstoy}, E., {Hill}, V., \& {Tosi}, M. 2009, ARA\&A, 47, 371

\bibitem[{{Urbaneja} {et~al.}(2008){Urbaneja}, {Kudritzki}, {Bresolin},
  {Przybilla}, {Gieren}, \& {Pietrzy{\'n}ski}}]{urbaneja2008}
{Urbaneja}, M.~A., {Kudritzki}, R.-P., {Bresolin}, F., {et~al.} 2008, ApJ,
  684, 118

\bibitem[{{Vogel} {et~al.}(1988){Vogel}, {Kulkarni}, \& {Scoville}}]{vogel1988}
{Vogel}, S.~N., {Kulkarni}, S.~R., \& {Scoville}, N.~Z. 1988, Nature, 334, 402

\bibitem[{{Weisz} {et~al.}(2011){Weisz}, {Dalcanton}, {Williams}, {Gilbert},
  {Skillman}, {Seth}, {Dolphin}, {McQuinn}, {Gogarten}, {Holtzman}, {Rosema},
  {Cole}, {Karachentsev}, \& {Zaritsky}}]{weisz2011}
{Weisz}, D.~R., {Dalcanton}, J.~J., {Williams}, B.~F., {et~al.} 2011, ApJ,
  739, 5

\bibitem[{{Yun}(1999)}]{yun1999}
{Yun}, M.~S. 1999, in IAU Symposium, Vol. 186, Galaxy Interactions at Low and
  High Redshift, ed. J.~E. {Barnes} \& D.~B. {Sanders}, 81

\end{thebibliography}
\end{document}